\documentclass[12pt]{article}

\usepackage[totalwidth=460truept,totalheight=613truept]{geometry}
\usepackage{latexsym,graphicx,amsfonts,amssymb,amsmath,xcolor}
\usepackage[hypertexnames=false,hidelinks]{hyperref}

\def\theequation{\arabic{section}.\arabic{equation}}

\renewcommand{\theequation}{\thesection.\arabic{equation}}
\global\arraycolsep=1truept
\renewcommand{\theequation}{\arabic{section}.\arabic{equation}}

\begin{document}

\null

\vskip1truecm

\begin{center}
{\huge \textbf{Dressed Propagators,}}

\vskip.8truecm

{\huge \textbf{Fakeon Self-Energy}}

\vskip.8truecm

{\huge \textbf{and Peak Uncertainty}}

\vskip1truecm

\textsl{Damiano Anselmi}

\vskip .1truecm

\textit{Dipartimento di Fisica \textquotedblleft Enrico
Fermi\textquotedblright , Universit\`{a} di Pisa, \\[0pt]
Largo B. Pontecorvo 3, 56127 Pisa, Italy}

\textit{INFN, Sezione di Pisa, Largo B. Pontecorvo 3, 56127 Pisa, Italy}

damiano.anselmi@unipi.it

\vskip1.5truecm

\textbf{Abstract}
\end{center}

We study the resummation of self-energy diagrams into dressed propagators in
the case of purely virtual particles and compare the results with those
obtained for physical particles and ghosts. The three geometric series
differ by infinitely many contact terms, which do not admit well-defined
sums. The peak region, which is outside the convergence domain, can only be
reached in the case of physical particles, thanks to analyticity. In the
other cases, nonperturbative effects become important. To clarify the
matter, we introduce the energy resolution $\Delta E$ around the peak and
argue that a \textquotedblleft peak uncertainty\textquotedblright\ $\Delta
E\gtrsim \Delta E_{\text{min}}\simeq \Gamma _{\text{f}}/2$ around energies $%
E\simeq m_{\text{f}}$ expresses the impossibility to approach the fakeon too
closely, $m_{\text{f}}$ being the fakeon mass and $\Gamma _{\text{f}}$ being
the fakeon width. The introduction of $\Delta E$ is also crucial to explain
the observation of unstable long-lived particles, like the muon. Indeed, by
the common energy-time uncertainty relation, such particles are also
affected by ill-defined sums at $\Delta E=0$, whenever we separate their
observation from the observation of their decay products. We study the
regime of large $\Gamma _{\text{f}}$, which applies to collider physics (and
situations like the one of the $Z$ boson), and the regime of small $\Gamma _{%
\text{f}}$, which applies to quantum gravity (and situations like the one of
the muon).

\vfill\eject

\section{Introduction}

\label{intro}\setcounter{equation}{0}

Perturbative quantum field theory is the most successful framework for the
investigation of the fundamental interactions of nature. Its predictions
have been repeatedly confirmed in the context of the standard model of
particle physics. Moreover, it has the chance to explain quantum gravity.

The principles on which it is based, which are locality, renormalizability
and unitarity,\ can be phrased in simple terms. Even unitarity, usually
formulated in terms of cut diagrams \cite%
{cutkosky,veltman,thooft,diagrammar,diagrammatica}, can be understood as a
collection of relatively simple algebraic identities \cite{diagrammarMio}.

The nonperturbative sector of quantum field theory is still elusive. Most
knowledge available today beyond the perturbation expansion, concerning the
anomalies, the running couplings and the particle widths, comes from the
resummation of the perturbative series, combined with analyticity.

The particle widths are obtained by resumming the geometric series due to
the self-energy diagrams. This is normally a straightforward operation and
gives the so-called dressed propagators. Yet, a geometric series has a
finite convergence radius. Since the \textquotedblleft peak
region\textquotedblright\ can only be reached by means of analyticity, the
issue must be reconsidered when analyticity is not available.

In this paper we investigate the dressed propagators of physical particles,
purely virtual particles and ghosts. We show that they differ by infinitely
many contact terms, which do not admit well-defined sums in the sense of
mathematical distributions. Unexpected properties emerge even in the case of
physical particles, concerning the experimental observation of long-lived,
unstable particles, like the muon.

Normally, the muon is treated as an approximately stable particle, due to
its long lifetime. Because its width is very small (around 10$^{-19}$GeV),
the free muon propagator is used, instead of the dressed one. So doing,
satisfactory predictions are obtained in most cases.

Nothing prevents us from using the dressed propagator of the muon, at least
in principle. Then we discover that the resummation of self-energies is
problematic even in the case of physical particles, when we want to separate
their observation from the observation of their decay products. If we make
the separation \textit{after} the resummation, we find a well-defined
result, which however vanishes: the theory predicts... no muon observation
at all! This result is, strictly speaking, correct, since the theory of
scattering deals with processes that occur between incoming\ states at $%
t=-\infty $ and outgoing\ states at $t=+\infty $: no matter how long the
muons live, they cannot survive till the end of time. Yet, it is troubling
that the dressed propagator fails to explain a simple phenomenon like the
observation of the muon.

Recapitulating, in order to explain the experimental fact that we do observe
the muon, we tweak the theory by pretending that the muon is a stable
particle, ignore the fact that the resummation of the self-energies kills
the possibility to observe it and use the free muon propagator. This
situation is clearly unsatisfactory. We call it \textquotedblleft the
problem of the muon\textquotedblright .

We show that the problem is solved by introducing the energy resolution $%
\Delta E$ around the peak, which is necessary to describe scattering
processes occurring within a finite amount of time $\Delta t<\infty $. Then
we obtain the correct result (i.e., that the muon is indeed observable) even
after the resummation of the self-energies into the dressed propagator.

The reason why a nonzero $\Delta E$ is necessary to observe the muon is that 
$\Delta E=0$ implies an infinite uncertainty on the measurement of time,
which makes every unstable particle decay before we can actually observe it.
It is impossible to observe an unstable particle with infinite resolving
power on the energy: because the observation requires a finite time
resolution $\bar{\Delta}t$ (smaller than the particle lifetime), the
energy-time uncertainty principle would be violated. Quantum field theory
\textquotedblleft knows it\textquotedblright\ and creates problems in the
crucial places.

\bigskip

The energy resolution $\Delta E$ is useful for many other purposes. In
particular, it allows us to clarify what happens in the cases of purely
virtual particles and ghosts, by properly treating the contact terms
mentioned above. We show that in those cases the geometric series cannot be
resummed in the peak region, because analyticity is unable to justify that
operation. We argue that nonperturbative effects play crucial roles there
and that, in the case of fakeons, the physical meaning is a new type of
uncertainty, which we call \textquotedblleft peak
uncertainty\textquotedblright . It codifies the reaction of the fakeon when
we attempt to \textquotedblleft approach it too closely\textquotedblright .

Basically, the energy resolution $\Delta E$ around the peak, in the
particle's rest frame, cannot be smaller than a certain minimum amount: $%
\Delta E\gtrsim \Delta E_{\text{min}}\simeq \Gamma _{\text{f}}/2$ around
energies $E\simeq m_{\text{f}}$, where $m_{\text{f}}$ and $\Gamma _{\text{f}%
} $ are the fakeon mass and the fakeon width, respectively. The uncertainty
relation actually provides the physical meaning of the \textquotedblleft
width\textquotedblright\ of a purely virtual particle.

We compare two phenomenological regimes: the regime of large $\Gamma _{\text{%
f}}$, which applies to collider physics and situations like the one of the $%
Z $ boson ($\Delta E\ll \Gamma _{\text{f}}/2$); and the regime of small $%
\Gamma _{\text{f}} $, which applies to quantum gravity and situations like
the one of the muon ($\Delta E\gg \Gamma _{\text{f}}/2$). The peak
uncertainty does not have a relation with the violation of microcausality 
\cite{causalityQG}, also due to fakeons. Ghosts are interested by a peak
uncertainty as well, but we cannot attach a physical significance to it.

\bigskip

Purely virtual particles, or fake particles, or \textquotedblleft
fakeons\textquotedblright \cite{fakeons}, can be used to propose new physics
beyond the standard model, by evading common constraints in collider physics 
\cite{Tallinn1}, offering new possibilities of solving discrepancies with
data \cite{Tallinn2}, or formulating a consistent theory of quantum gravity 
\cite{LWgrav}, which is experimentally testable due to its predictions in
inflationary cosmology \cite{ABP}. By pursuing an approach that is radically
different from the previous ones, fakeons define a new diagrammatics \cite%
{diagrammarMio}, which can be implemented in softwares like FeynCalc,
FormCalc, LoopTools and Package-X \cite{calc} and used to work out physical
predictions. The only requirement is that fakeons be massive and non
tachyonic (the real part of the squared mass should be strictly positive).
The no tachyon condition is especially important in cosmology and leads to
the ABP\ bound \cite{ABP}, which is crucial for the sharp prediction of the
tensor-to-scalar ratio $r$. %It is also possible \cite%
%{LWformulation,LWunitarity,FakeonsLW} to avoid certain troubles of the
%Lee-Wick models \cite{leewick,LWQED,lee,nakanishi,CLOP,grinstein} by
%switching to theories of particles and fakeons. 
For proofs to all orders, see \cite{fakeons,diagrammarMio}.

The investigation of dressed propagators is relevant to model building for
new physics beyond the standard model, which may involve relatively light
fake particles. The comparison with physical particles and ghosts is useful
to better appreciate the key points. The ghost prescription we treat in this
paper is the usual Feynman $i\epsilon $ one, which gives positive energy,
but indefinite metric. Equivalently, it is the one inherited from the Wick
rotation of the Euclidean theory. It is commonly used for the Pauli-Villars
fields \cite{PV}, the Faddeev-Popov ghosts and the longitudinal and temporal
components of the gauge fields. However, the resummation of self-energies in
the regularization and gauge-fixing sectors of a gauge theory is not
strictly necessary, which probably explains why it has not been investigated
extensively so far. In higher-derivative gravity, the same prescription
defines the Stelle theory \cite{stelle} (see also \cite{agravity}). There,
the violation of unitary makes the study of dressed propagators less
compelling. Other ghost prescriptions, like the Lee-Wick one \cite%
{leewick,LWQED,lee,nakanishi,CLOP,grinstein}, are discussed in a separate
paper \cite{FakeonsLW}.

The paper is organized as follows. In section \ref{formal} we formally resum
the self-energy diagrams in the three cases. In section \ref{differences} we
show that the expansions differ by infinitely many contact terms, which do
not sum into well-defined mathematical distributions. In section \ref%
{physical} we study the dressed propagators of physical particles and show
that, although analyticity allows us to trust the resummation everywhere,
including the peaks, the result cannot explain the observation of the muon.
We solve this problem by introducing the energy resolution $\Delta E$. In
section \ref{fakeons} we study purely virtual particles and show that the
resummation can be trusted only if $\Delta E$ satisfies a certain bound,
which we interpret as a peak uncertainty relation. In section \ref{ghosts}
we show that the resummation faces similar, but different challenges in the
case of ghosts. In section \ref{phenom} we study the regimes $\Delta E\ll
\Gamma/2 $ and $\Delta E\gg \Gamma/2 $ and provide phenomenological
candidates for the dressed propagators around the peaks. In section \ref%
{microc} we compare the peak uncertainty relation with the violation of
microcausality and show that the two properties are essentially unrelated.
Section \ref{conclusions} contains the conclusions. The appendices collect
technical derivations and proofs of formulas and properties used in the
paper.

\section{Formal resummation of self energies}

\label{formal}\setcounter{equation}{0} 
\begin{figure}[t]
\begin{center}
\includegraphics[width=16truecm]{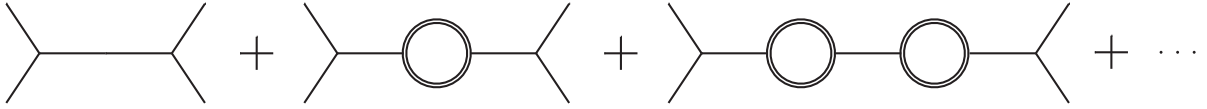}
\end{center}
\caption{Resummation of self-energies}
\label{ABC}
\end{figure}

In this section we formally resum the self-energy diagrams (see fig. \ref%
{ABC}) in the cases of physical particles $\varphi $, fakeons $\chi $ and
ghosts $\phi $. We postpone the discussion about the domains where the
resummations make sense to the next sections.

\subsection{Physical particles}

The tree-level propagator of a physical particle $\varphi $ is%
\begin{equation}
P_{\varphi }(p^{2},m^{2},\epsilon )=\frac{i}{p^{2}-m^{2}+i\epsilon }.
\label{propag}
\end{equation}%
Let $(-i)\Sigma (p^{2})$ denote the amputated one-particle irreducible (1PI) 
$\varphi $ two-point function, which has the form%
\begin{equation*}
(-i)\Sigma (p^{2})=-\sum_{j}\theta (p^{2}-M_{j}^{2})B_{j}(p^{2})-iA(p^{2}),
\end{equation*}%
where $A(p^{2})$ and $B_{j}(p^{2})$ are real functions and $M_{j}^{2}$ are
the thresholds of the $\varphi $ decay processes. The sum is nonnegative due
to the optical theorem, which gives Im$[-\Sigma ]\geqslant 0$.

The full $\varphi $ two-point function $\hat{P}_{\varphi }$, obtained by
resumming a geometric series, is%
\begin{equation}
\hat{P}_{\varphi }=P_{\varphi }\sum_{n=0}^{\infty }(-i\Sigma P_{\varphi
})^{n}=\frac{i}{p^{2}-m^{2}-\Sigma (p^{2})+i\epsilon }.  \label{propago}
\end{equation}%
For most purposes, it is sufficient to study (\ref{propago}) around the
peak. Assuming that $\Sigma (p^{2})$ is nonzero there, we can use the
approximation 
\begin{equation}
\Sigma (p^{2})\simeq Z^{-1}(\Delta m^{2}-im_{\text{ph}}\Gamma
)+(1-Z^{-1})(p^{2}-m^{2}),\qquad m_{\text{ph}}^{2}\equiv m^{2}+\Delta
m^{2},\qquad \Gamma \geqslant 0,  \label{approx}
\end{equation}%
with $Z$ real and positive\footnote{%
The approximation amounts to consider the Taylor expansion of $\Sigma
(p^{2}) $ around $p^{2}=m^{2}$ and neglect $\mathcal{O}((p^{2}-m^{2})^{2})$
in the real part, $\mathcal{O}(p^{2}-m^{2})$ in the imaginary part. The
optical theorem Im$[-\Sigma ]\geqslant 0$ implies $\Gamma \geqslant 0$. We
cannot include further corrections to the imaginary part, because the
optical theorem would then force us to include them all.}. We obtain%
\begin{equation}
\hat{P}_{\varphi }\simeq \frac{iZ}{p^{2}-m_{\text{ph}}^{2}+i(\tilde{\epsilon}%
+m_{\text{ph}}\Gamma )}  \label{propaga}
\end{equation}%
where $\tilde{\epsilon}=\epsilon Z$.

\subsection{Purely virtual particles}

\label{purely}

Purely virtual particles $\chi $ can be obtained either from physical
particles or ghosts, by changing their quantization prescriptions into the
fakeon ones. We first concentrate on the fakeons obtained from physical
particles. Their tree-level propagator is \cite{causalityQG}%
\begin{equation}
P_{\chi }(p^{2},m^{2},\epsilon )=\frac{i(p^{2}-m^{2})}{(p^{2}-m^{2})^{2}+%
\epsilon ^{2}}\underset{\epsilon \rightarrow 0}{\rightarrow }\mathcal{P}%
\frac{i}{p^{2}-m^{2}},  \label{pQtree}
\end{equation}%
where $\mathcal{P}$ denotes the Cauchy principal value. Only minor
modifications, discussed later, are required to treat the fakeons obtained
from ghosts, whose free propagators are the opposite of (\ref{pQtree}). 
\begin{figure}[t]
\begin{center}
\includegraphics[width=6truecm]{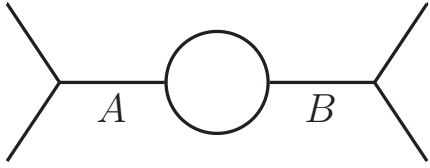}
\end{center}
\caption{Internal lines that, if broken, disconnect the diagram}
\label{AB}
\end{figure}

Formula (\ref{pQtree}) applies to any line that, if broken, disconnects the
diagram. Examples are the lines A or B in fig. \ref{AB}. It cannot be used
inside the loop diagrams, as shown in ref. \cite{wheelerons} (see also \cite%
{diagrammarMio}), but this is not of our concern in the present paper.

If $(-i)\Sigma (p^{2})$ denotes the amputated 1PI $\chi $ two-point
function, the formal resummation gives the dressed propagator 
\begin{equation}
\hat{P}_{\chi }=P_{\chi }\sum_{n=0}^{\infty }(-i\Sigma P_{\chi })^{n}=\frac{%
i(p^{2}-m^{2})}{(p^{2}-m^{2})[p^{2}-m^{2}-\Sigma (p^{2})]+\epsilon ^{2}}.
\label{sumfake}
\end{equation}%
When we do not specify the arguments of $P_{\chi }$, we mean $P_{\chi
}(p^{2},m^{2},\epsilon )$.

If we study the approximation\ (\ref{approx}) around the peak, we find that $%
\Gamma $ is nonnegative again, by the optical theorem. It receives
contributions from the physical particles circulating in the 1PI $\chi $
two-point function. The formal result is thus%
\begin{equation}
\hat{P}_{\chi }\simeq \frac{iZ(p^{2}-m^{2})}{(p^{2}-m^{2})(p^{2}-m_{\text{ph}%
}^{2}+im_{\text{ph}}\Gamma )+\tilde{\epsilon}^{2}}.  \label{phatf}
\end{equation}%
where $\tilde{\epsilon}=\epsilon Z^{1/2}$.

There are actually other options to perform the resummation in the case of
purely virtual particles. Since we cannot predict the final outcome in the
peak region, which, as we have anticipated, triggers nonperturbative
effects, it is interesting to explore the alternatives. In particular, the
powers of $\Delta m^{2}$ redefine the squared mass, which may or may not
have an impact on the problems we consider here. Formula (\ref{phatf})
describes the situation where they do. If we assume that they do not, we can
proceed as explained in appendix \ref{alternative}, where we resum $\Delta
m^{2}$ by default into $m_{\text{ph}}^{2}$. The result is%
\begin{equation}
\hat{P}_{\chi }\simeq \frac{iZ(p^{2}-m_{\text{ph}}^{2})}{(p^{2}-m_{\text{ph}%
}^{2})(p^{2}-m_{\text{ph}}^{2}+im_{\text{ph}}\Gamma )+\tilde{\epsilon}^{2}}.
\label{phatf2}
\end{equation}

Summarizing, we have two candidates for the dressed propagator of purely
virtual particles: (\ref{phatf}) and (\ref{phatf2}). Neither is
satisfactory, though. We can see it by noting that, in the limit $\tilde{%
\epsilon}\rightarrow 0$, both (\ref{phatf}) and (\ref{phatf2}) give the same
result we obtain from the physical particle, formula (\ref{propaga}). The
difference between a fakeon and a physical particle appears to be washed
away by the resummation. This cannot be correct, since if we expand the sums
we must find back what we summed.

Something suspicious is the simplification of $p^{2}-m^{2}$ in (\ref{phatf})
and $p^{2}-m_{\text{ph}}^{2}$ (\ref{phatf2}) between the numerator and the
denominator when $\tilde{\epsilon}$ tends to zero. The point is that we
cannot take the limit $\tilde{\epsilon}\rightarrow 0$ so naively.

\subsection{Ghosts}

Ghosts $\phi $ have the tree-level propagator%
\begin{equation}
P_{\phi }(p^{2},m^{2},\epsilon )=-\frac{i}{p^{2}-m^{2}+i\epsilon }.
\label{propagh}
\end{equation}%
Although the prescription is the Feynman one, the minus sign in front makes
a crucial difference, as we are going to show. If, as before, $(-i)\Sigma
(p^{2})$ denotes the amputated 1PI $\phi $ two-point function, the
resummation gives%
\begin{equation*}
\hat{P}_{\phi }=P_{\phi }\sum_{n=0}^{\infty }(-i\Sigma P_{\phi })^{n}=-\frac{%
i}{p^{2}-m^{2}+\Sigma (p^{2})+i\epsilon }.
\end{equation*}%
For convenience, we rearrange the approximation (\ref{approx}) around $%
p^{2}\simeq m^{2}$ as%
\begin{equation}
\Sigma (p^{2})\simeq Z^{-1}(-\Delta m^{2}-im_{\text{ph}}\Gamma
)+(Z^{-1}-1)(p^{2}-m^{2}),\qquad m_{\text{ph}}^{2}\equiv m^{2}+\Delta
m^{2},\qquad \Gamma \geqslant 0,  \label{approxgh}
\end{equation}%
by changing the conventions for $\Delta m^{2}$ and $Z$. We can still assume $%
\Gamma \geqslant 0$, because, if $\phi $ \textquotedblleft
decays\textquotedblright\ into lighter physical particles, $-i\Sigma $ is
the same as before at one loop around $p^{2}\simeq m^{2}$. We obtain%
\begin{equation}
\hat{P}_{\phi }\simeq -\frac{iZ}{p^{2}-m_{\text{ph}}^{2}+i(\tilde{\epsilon}%
-m_{\text{ph}}\Gamma )}  \label{resugh}
\end{equation}%
close to the ghost peak, where $\tilde{\epsilon}=\epsilon Z$. Note\ the
crucial sign difference appearing in $\tilde{\epsilon}-m_{\text{ph}}\Gamma $
with respect to (\ref{propaga}). Again, we cannot take the limit $\tilde{%
\epsilon}\rightarrow 0$ naively, because it reverses the prescription.

\section{Differences among physical particles, fake particles and ghosts}

\label{differences}\setcounter{equation}{0}

In this section we analyze of the differences among the perturbative
expansions of $\hat{P}_{\varphi }$, $\hat{P}_{\chi }$ and $\hat{P}_{\phi }$.
We assume $Z=1$ and $\Delta m^{2}=0$, to simplify the formulas and
concentrate on the effects of the width $\Gamma $. We also assume that $m$
and $\Gamma $ are the same in the three cases. We have 
\begin{eqnarray}
\hat{P}_{\varphi } &\simeq &\frac{i}{p^{2}-m^{2}+i(\epsilon +m\Gamma )}%
,\qquad \hat{P}_{\chi }\simeq \frac{i(p^{2}-m^{2})}{%
(p^{2}-m^{2})(p^{2}-m^{2}+im\Gamma )+\epsilon ^{2}},  \notag \\
\hat{P}_{\phi } &\simeq &-\frac{i}{p^{2}-m^{2}+i(\epsilon -m\Gamma )},
\label{p2}
\end{eqnarray}%
respectively.

Since $\epsilon $ is a mathematical artifact, it must tend to zero at the
end. We can take two attitudes towards this limit:\ we can study the limit $%
\epsilon \rightarrow 0$ term by term in the perturbative expansion and then
resum; or first resum and let $\epsilon \rightarrow 0$ at the end. The
former is what perturbative quantum field theory asks us to do, strictly
speaking, so for the time being we concentrate on this option.

The formulas (\ref{p2}) have to be meant as shorthand expressions for their
perturbative expansions, which are the expansion in powers of $\Gamma $. We
show that, in the cases of purely virtual particles and ghosts, the
resummations are not legitimate, because they do not make sense as
mathematical distributions. We mostly concentrate on the real parts, but the
arguments and results easily extend to the whole $\hat{P}_{\varphi }$, $\hat{%
P}_{\chi }$ and $\hat{P}_{\phi }$.

\subsection{Difference with ghosts}

It is convenient to start from the difference Im$[im^{2}(\hat{P}_{\varphi }-%
\hat{P}_{\phi })]$ between physical particles and ghosts, which can be
written as%
\begin{equation}
\text{Im}[im^{2}(\hat{P}_{\varphi }-\hat{P}_{\phi })]=\frac{i}{2}\left[ 
\frac{1}{x+i\hat{\Gamma}+i\hat{\epsilon}}-\frac{1}{x+i\hat{\Gamma}-i\hat{%
\epsilon}}+\frac{1}{x-i\hat{\Gamma}+i\hat{\epsilon}}-\frac{1}{x-i\hat{\Gamma}%
-i\hat{\epsilon}}\right] ,  \label{diffa}
\end{equation}%
where $x\equiv (p^{2}-m^{2})/m^{2}$, $\hat{\epsilon}=\epsilon /m^{2}$ and $%
\hat{\Gamma}=\Gamma /m$. Naively, the right-hand side tends to zero when $%
\hat{\epsilon}$ tends to zero. Considering $\pm i\hat{\Gamma}$ as shifts of $%
x$, and recalling that we are working on the series expansions in powers of $%
\hat{\Gamma}$, we can formally write%
\begin{equation}
\left. \text{Im}[im^{2}(\hat{P}_{\varphi }-\hat{P}_{\phi })]\right\vert _{%
\hat{\epsilon}\rightarrow 0}=\pi \left[ \delta (x+i\hat{\Gamma})+\delta (x-i%
\hat{\Gamma})\right] =2\pi \sum_{n=0}^{\infty }\frac{(-\hat{\Gamma}^{2})^{n}%
}{(2n)!}\delta ^{(2n)}(x).  \label{diffgh}
\end{equation}%
The expressions $\delta (x\pm i\hat{\Gamma})$ are, again, shorthand
notations for their perturbative expansions in powers of $\hat{\Gamma}$. We
see that the difference (\ref{diffa})\ is not zero, but an infinite series
of contact terms. Now, Im$[im^{2}\hat{P}_{\varphi }]$ resums into a
well-behaved function, which is the Breit-Wigner formula%
\begin{equation*}
\text{Im}[im^{2}\hat{P}_{\varphi }]=\frac{\hat{\epsilon}+\hat{\Gamma}}{%
x^{2}+(\hat{\epsilon}+\hat{\Gamma})^{2}}\underset{\hat{\epsilon}\rightarrow 0%
}{\longrightarrow }\frac{\hat{\Gamma}}{x^{2}+\hat{\Gamma}^{2}}.
\end{equation*}%
We conclude that $\left. \text{Im}[im^{2}\hat{P}_{\phi }]\right\vert _{\hat{%
\epsilon}\rightarrow 0}$ is equal to the same thing minus $2\pi \Delta _{%
\hat{\Gamma}}(x)$, where%
\begin{equation}
\Delta _{\hat{\Gamma}}(x)\equiv \sum_{n=0}^{\infty }\frac{(-\hat{\Gamma}%
^{2})^{n}}{(2n)!}\delta ^{(2n)}(x)=\frac{1}{2}[\delta (x+i\hat{\Gamma}%
)+\delta (x-i\hat{\Gamma})].  \label{dgn}
\end{equation}

The question is: what is $\Delta _{\hat{\Gamma}}(x)$? Evidently, it is not a
function. We could accept it as a mathematical distribution. In appendix \ref%
{singulard} we show in detail that it is not even a distribution. Here we
discuss the physical meaning of this fact.

At the practical level, we cannot determine the momenta of the incoming
particles with absolute precision, so the true incoming\ state can be
written as a superposition%
\begin{equation*}
|\text{in}\rangle =\int \frac{\mathrm{d}^{3}\mathbf{k}_{1}}{2k_{1}^{0}(2\pi
)^{3}}\frac{\mathrm{d}^{3}\mathbf{k}_{2}}{2k_{2}^{0}(2\pi )^{3}}f_{1}(%
\mathbf{k}_{1})f_{2}(\mathbf{k}_{2})|\mathbf{k}_{1},\mathbf{k}_{2},\text{in}%
\rangle ,
\end{equation*}%
where $f_{1}(\mathbf{k}_{1})$ and $f_{2}(\mathbf{k}_{2})$ are the amplitudes
of the wave packets and $|\mathbf{k}_{1},\mathbf{k}_{2},$in$\rangle $
denotes the ideal state where the incoming particles have definite momenta $%
\mathbf{k}_{1}$ and $\mathbf{k}_{2}$. Mathematically, the wave packets can
be seen as test functions and the correlation functions as distributions.

If we choose the wave packets in a suitable way, $\Delta _{\hat{\Gamma}}(x)$
originates absurdities of new types (not directly related the violations of
unitarity, typical of ghosts). For example, if we take the convolution of
the function $\Delta _{\hat{\Gamma}}(x)$ with certain wave packets of
incoming momenta, we can generate a pole at $p^{2}=m^{2}$ out of nowhere,
with no width: 
\begin{equation*}
\int_{-\infty }^{+\infty }\mathrm{d}y\hspace{0.01in}\Delta _{\hat{\Gamma}%
}(x-y)\frac{y}{y^{2}+\hat{\Gamma}^{2}}=\frac{1}{x}\frac{x^{2}+2\hat{\Gamma}%
^{2}}{x^{2}+4\hat{\Gamma}^{2}}=\frac{m^{2}}{p^{2}-m^{2}}\frac{%
(p^{2}-m^{2})^{2}+2m^{2}\Gamma ^{2}}{(p^{2}-m^{2})^{2}+4m^{2}\Gamma ^{2}}.
\end{equation*}

%These facts may point to a fundamental problem or mean that the resummation
%is illegitimate. We opt for the second possibility, because a theory with
%ghosts may be physically unacceptable, but should make sense mathematically.

\subsection{Perturbative expansions by means of distributions}

To extend this analysis to purely virtual particles and study $\hat{P}%
_{\varphi }$, $\hat{P}_{\chi }$ and $\hat{P}_{\phi }$ further, it is useful
to work out a systematic expansion by means of distributions. For example,
if we use identities such as%
\begin{equation*}
\frac{\mathrm{d}}{\mathrm{d}\hat{\epsilon}}\frac{\hat{\epsilon}}{x^{2}+\hat{%
\epsilon}^{2}}=-\frac{\mathrm{d}}{\mathrm{d}x}\frac{x}{x^{2}+\hat{\epsilon}%
^{2}},\qquad \frac{\mathrm{d}}{\mathrm{d}\hat{\epsilon}}\frac{x}{x^{2}+\hat{%
\epsilon}^{2}}=\frac{\mathrm{d}}{\mathrm{d}x}\frac{\hat{\epsilon}}{x^{2}+%
\hat{\epsilon}^{2}},
\end{equation*}%
the expansion of the Breit-Wigner formula is immediately obtained:%
\begin{equation}
\text{Im}[im^{2}\hat{P}_{\varphi }]=\frac{\hat{\epsilon}+\hat{\Gamma}}{%
x^{2}+(\hat{\epsilon}+\hat{\Gamma})^{2}}=\sum_{n=0}^{\infty }\frac{\hat{%
\Gamma}^{n}}{n!}\frac{\mathrm{d}^{n}}{\mathrm{d}\hat{\epsilon}^{n}}\frac{%
\hat{\epsilon}}{x^{2}+\hat{\epsilon}^{2}}\rightarrow \pi \Delta _{\hat{\Gamma%
}}(x)+\frac{1}{\hat{\Gamma}}\sum_{n=0}^{\infty }\frac{(-\hat{\Gamma}%
^{2})^{n+1}}{(2n+1)!}\mathcal{P}^{(2n+1)}\frac{1}{x},  \label{physum}
\end{equation}%
where $\mathcal{P}^{(k)}$ denotes the $k$-th derivative of the Cauchy
principal value. We have taken $\hat{\epsilon}$ to zero term by term in the
last expression.

In the case of ghosts, the same procedure gives%
\begin{equation}
\text{Im}[im^{2}\hat{P}_{\phi }]=\frac{\hat{\Gamma}-\hat{\epsilon}}{x^{2}+(%
\hat{\Gamma}-\hat{\epsilon})^{2}}\rightarrow -\pi \Delta _{\hat{\Gamma}}(x)+%
\frac{1}{\hat{\Gamma}}\sum_{n=0}^{\infty }\frac{(-\hat{\Gamma}^{2})^{n+1}}{%
(2n+1)!}\mathcal{P}^{(2n+1)}\frac{1}{x}.  \label{gho}
\end{equation}%
The derivatives of the principal values disappear in the difference Im$%
[im^{2}(\hat{P}_{\varphi }-\hat{P}_{\phi })]$, which confirms (\ref{diffgh}%
). Thus, the powers of $\Gamma $ cannot be resummed at $\hat{\epsilon}=0$ in
the case of ghosts.

\subsection{Difference with purely virtual particles}

\label{anyway}

Now we switch to fakeons. If we want to resum the powers of $\Gamma $ at $%
\hat{\epsilon}=0$, we must deal with 
\begin{equation*}
\text{Im}[im^{2}\hat{P}_{\chi }]=\frac{x^{2}\hat{\Gamma}}{(x^{2}+\hat{%
\epsilon}^{2})^{2}+x^{2}\hat{\Gamma}^{2}}=-\frac{1}{\hat{\Gamma}}%
\sum_{n=0}^{\infty }\frac{(-x^{2}\hat{\Gamma}^{2})^{n+1}}{(x^{2}+\hat{%
\epsilon}^{2})^{2n+2}}.
\end{equation*}%
Taking $\hat{\epsilon}\rightarrow 0$ term by term in the last expression, we
obtain powers of the Cauchy principal value, which can be treated by means
of the coincidence-splitting method explained in appendix \ref{split},
formula (\ref{csplit}), Thus, 
\begin{equation}
\text{Im}[im^{2}\hat{P}_{\chi }]\rightarrow \frac{1}{\hat{\Gamma}}%
\sum_{n=0}^{\infty }\frac{(-\hat{\Gamma}^{2})^{n+1}}{(2n+1)!}\mathcal{P}%
^{(2n+1)}\frac{1}{x}.  \label{cspli}
\end{equation}%
The difference with respect to the physical particle is one half of (\ref%
{diffgh}): 
\begin{equation*}
\text{Im}[im^{2}(\hat{P}_{\varphi }-\hat{P}_{\chi })]\rightarrow \pi \Delta
_{\hat{\Gamma}}(x).
\end{equation*}

It is clear that only the physical combination (\ref{physum}) is well
defined, while (\ref{gho}) and (\ref{cspli}) are not, since $\Delta _{\hat{%
\Gamma}}(x)$ is not a distribution. The real parts of $im^{2}\hat{P}%
_{\varphi ,\chi ,\phi }$ can be studied similarly and lead to analogous
conclusions.

The perturbative expansion is missing something. On the one hand, the
differences proportional to $\Delta _{\hat{\Gamma}}(x)$ cannot be ignored.
On the other hand, they do not describe what is missing accurately enough,
due to the (mathematical and physical) difficulties of $\Delta _{\hat{\Gamma}%
}(x)$. As we are going to show in the next sections, the problems become
nonperturbative around the peaks of purely virtual particles and ghosts.

\section{Dressed propagator of physical particles}

\label{physical}\setcounter{equation}{0}

In this section we discuss the convergence of the formal resummation (\ref%
{propaga}) in the case of physical particles $\varphi $, pointing out
several nontrivial facts, to prepare the discussion of the next section
about purely virtual particles. We assume that $\Gamma $, $\Delta m^{2}$ and 
$Z-1$ are generic, of order $\lambda ^{2}$ in some (small) dimensionless
coupling $\lambda $.

The condition of convergence for the resummation (\ref{propago}) is (keeping 
$\epsilon $ for future use)%
\begin{equation}
\frac{|\Sigma (p^{2})|}{|p^{2}-m^{2}+i\epsilon |}<1.  \label{conver}
\end{equation}%
It can be refined by combining it with the approximation (\ref{approx}) and
using (\ref{propaga}). For example, $Z$ just affects the overall constant,
so the corrections due to it can be resummed first with no difficulties.
Once the powers of $Z-1$ are resummed, we can focus on the resummation of
the powers of $\Gamma $ and $\Delta m^{2}$, which is convergent for 
\begin{equation}
\frac{(\Delta m^{2})^{2}+m_{\text{ph}}^{2}\Gamma ^{2}}{(p^{2}-m^{2})^{2}+%
\tilde{\epsilon}^{2}}<1,  \label{converga}
\end{equation}%
i.e., sufficiently away from the peak region, which is $p^{2}\simeq m_{\text{%
ph}}^{2}=m^{2}+\Delta m^{2}$. If we resum the powers of $\Delta m^{2}$ into $%
m_{\text{ph}}^{2}$ by default, and focus on the powers of $\Gamma $, then
the convergence condition reads 
\begin{equation}
\frac{m_{\text{ph}}^{2}\Gamma ^{2}}{(p^{2}-m_{\text{ph}}^{2})^{2}+\tilde{%
\epsilon}^{2}}<1.  \label{converga2}
\end{equation}

The validity of the resummation can be extended beyond the bounds just
found, thanks to analyticity.

A priori, we cannot be sure that analyticity holds after the resummation, so
we treat it as a work hypothesis. A second hypothesis is that essential
singularities, which are invisible at the level of the perturbative
expansion, do not contribute or are negligible. These assumptions can be
validated or rejected by experiments.

The correct why to phrase what is happening is as follows. Since

a) the Feynman prescription is analytic and

b) the denominator of (\ref{propaga}) never vanishes,

\noindent there is an analytic way to extend the result beyond the region of
convergence, that is to say, close to the peak. Moreover, since

c) experiments support the result obtained this way,

\noindent the resummed $\hat{P}_{\varphi }$ is correct and the resummation
holds everywhere.

\subsection{The problem of the muon and its solution}

\label{muonprobl}

The comparison between experiments and theoretical predictions requires
further analysis. Let us consider a process like $e^{+}e^{-}\rightarrow
e^{+}e^{-}$. The contribution $\mathcal{M}_{\varphi }$ of the diagrams of
fig. \ref{ABC} to the total amplitude $\mathcal{M}$ of the process is equal
to $i\hat{P}_{\varphi }$ (assuming that the external vertices are equal to $%
-i$). Typically, $\mathcal{M}_{\varphi }$ is the dominant contribution in a
neighborhood of $p^{2}=m_{\text{ph}}^{2}$. In what follows, we concentrate
on it.

\begin{figure}[t]
\begin{center}
\includegraphics[width=16truecm]{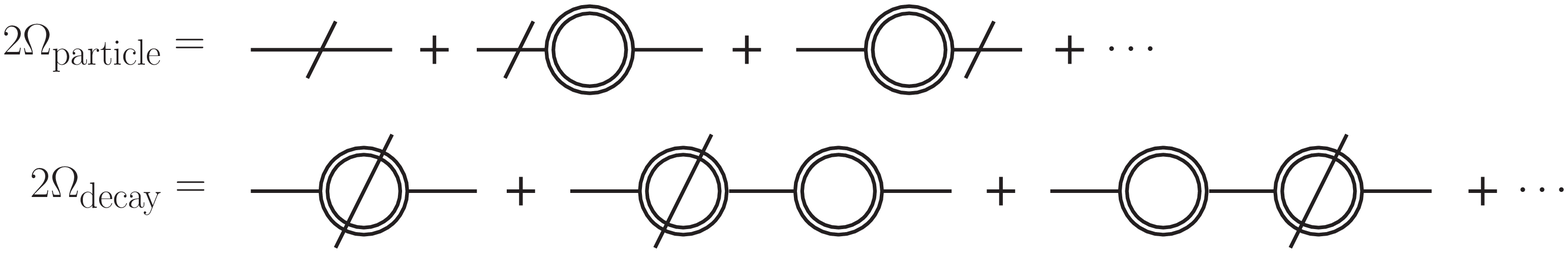}
\end{center}
\caption{Cut diagrams contributing to $2\Omega _{\text{particle}}$ and $%
2\Omega _{\text{decay}}$, respectively}
\label{cutdiagrams}
\end{figure}

By the optical theorem, the imaginary part Im$[2\mathcal{M}]$ is
proportional to the sum of the cross sections of the processes $%
e^{+}e^{-}\rightarrow $ $X$, where $X$ denotes any set of outgoing states.
The processes involved in 2Im$[\mathcal{M}_{\varphi }]$, which can be read
by cutting the diagrams as shown in fig. \ref{cutdiagrams}, are 
\begin{equation*}
e^{+}e^{-}\rightarrow \varphi ,\qquad e^{+}e^{-}\rightarrow \text{decay
products of }\varphi \text{.}
\end{equation*}%
The former is the process where the particle is physically observed before
it decays (as in the case $\varphi $ = muon). The latter is the process
where $\varphi $ is not observed directly (typically, because its lifetime
is too short, as in the case $\varphi $ = $Z$ boson): its decay products are
observed, instead. Thus, we have%
\begin{equation}
2\text{Im}[\mathcal{M}_{\varphi }]\simeq \int \mathrm{d}\Pi _{f}\hspace{%
0.01in}|\mathcal{M}_{e^{+}e^{-}\rightarrow \varphi }|^{2}+\int \mathrm{d}\Pi
_{f}\hspace{0.01in}|\mathcal{M}_{e^{+}e^{-}\rightarrow \text{decay products
of }\varphi }|^{2},  \label{opto}
\end{equation}%
around the $\varphi $ peak.

Diagrammatically, the two contributions correspond to the first and second
lines of fig. \ref{cutdiagrams}, which give the sums reported in formulas (%
\ref{omica4}) of appendix \ref{split}. The integrals on the phase spaces $%
\Pi _{f}$ of the outgoing states are originated by the cut propagators,
according to the rules of cut diagrams. The vertices and propagators that
lie to the right of the cut are the normal ones (as in $\mathcal{M}$), while
those that lie to the left of the cut are the complex conjugate ones (as in $%
\mathcal{M}^{\dagger }$). We are assuming that the energy flows from the
right to the left.

We have, from (\ref{propago}) and (\ref{omica4}),%
\begin{equation*}
2\text{Im}[\mathcal{M}_{\varphi }]=2\text{Re}[\hat{P}_{\varphi }]=\frac{%
2(\epsilon -\text{Im}[\Sigma ])}{|p^{2}-m^{2}-\Sigma +i\epsilon |^{2}}%
=2\Omega _{\varphi \hspace{0.01in}\text{particle}}+2\Omega _{\varphi \hspace{%
0.01in}\text{decay}},
\end{equation*}%
where 
\begin{eqnarray}
2\Omega _{\varphi \hspace{0.01in}\text{particle}} &=&\frac{2\epsilon }{%
|p^{2}-m^{2}-\Sigma +i\epsilon |^{2}}\underset{\epsilon \rightarrow 0}{%
\longrightarrow }\int \mathrm{d}\Pi _{f}\hspace{0.01in}|\mathcal{M}%
_{e^{+}e^{-}\rightarrow \varphi }|^{2},  \label{omica1} \\
2\Omega _{\varphi \hspace{0.01in}\text{decay}} &=&-\frac{2\text{Im}[\Sigma ]%
}{|p^{2}-m^{2}-\Sigma +i\epsilon |^{2}}\underset{\epsilon \rightarrow 0}{%
\longrightarrow }\int \mathrm{d}\Pi _{f}\hspace{0.01in}|\mathcal{M}%
_{e^{+}e^{-}\rightarrow \text{decay products of }\varphi }|^{2}.
\label{omica2}
\end{eqnarray}%

Using the approximation (\ref{approx}) around the peak, we find 
\begin{eqnarray}
\Omega _{\varphi \hspace{0.01in}\text{particle}} &\simeq &\frac{\tilde{%
\epsilon}Z}{(p^{2}-m_{\text{ph}}^{2})^{2}+(\tilde{\epsilon}+m_{\text{ph}%
}\Gamma )^{2}},  \label{Ms} \\
\Omega _{\varphi \hspace{0.01in}\text{decay}} &\simeq &\frac{m_{\text{ph}%
}Z\Gamma }{(p^{2}-m_{\text{ph}}^{2})^{2}+(\tilde{\epsilon}+m_{\text{ph}%
}\Gamma )^{2}}.  \label{Md}
\end{eqnarray}

Again, we can let $\epsilon $ tend to zero\ before or after the resummation.
Here we study the limit $\epsilon \rightarrow 0$ after the resummation,
below (and in appendix \ref{split}) we say more about the limit before the
resummation.

Taking $\epsilon $ to zero in (\ref{Ms}) and (\ref{Md}), we obtain%
\begin{equation}
\Omega _{\varphi \hspace{0.01in}\text{particle}}\rightarrow 0,\qquad \Omega
_{\varphi \hspace{0.01in}\text{decay}}\rightarrow \frac{m_{\text{ph}}Z\Gamma 
}{(p^{2}-m_{\text{ph}}^{2})^{2}+m_{\text{ph}}^{2}\Gamma ^{2}}.  \label{limpa}
\end{equation}%
The conclusion is that an unstable particle has zero probability of being
observed. Such a result is in contradiction with experiments, since the muon
is unstable, but can be observed before it decays: $\Omega _{\varphi \hspace{%
0.01in}\text{particle}}$ should not be zero; $\Omega _{\varphi \hspace{0.01in%
}\text{decay}}$ should vanish, instead. This is what we call
\textquotedblleft the problem of the muon\textquotedblright .

In quantum field theory a number of shortcuts are commonly adopted to
simplify the derivations of general formulas. In particular, the scattering
processes are usually meant to occur between incoming\ states at $t=-\infty $
and outgoing\ states at $t=+\infty $. Since the time interval $\Delta t$
separating them is, strictly speaking, infinite, every unstable particle has
enough time to decay before being observed, in agreement with the result $%
\Omega _{\varphi \hspace{0.01in}\text{particle}}=0$ obtained above. In this
sense, there is no contradiction. However, the observation of the muon is a
fact and we should be able to account for it.

In practical situations the scattering processes occur within some finite
time interval $\Delta t$, much larger than the duration $\bar{\Delta}t$ of
the interactions involved in the process. The prediction $\Omega _{\varphi 
\hspace{0.01in}\text{particle}}=0$ remains correct whenever $\Delta t$ is
also much larger than, say, the muon lifetime $\tau _{\mu }$. We have a
problem for $\bar{\Delta}t\ll \Delta t<\tau _{\mu }$, since the muon has not
enough time to decay in that case. After the resummation of the
self-energies into the dressed propagator, we still obtain the prediction $%
\Omega _{\varphi \hspace{0.01in}\text{particle}}=0$, which is in
contradiction with the phenomenon we observe.

In principle, we should undertake the task of rederiving all the basic
formulas of quantum field theory for scattering processes where incoming\
and outgoing\ states are separated by a finite $\Delta t$. Once we do so,
the predictions end up depending on $\Delta t$. They also depend on the
energy resolution $\Delta E\sim 1/\bar{\Delta}t$, where $\bar{\Delta}t$ is
the time uncertainty. Indeed, only $\Delta t=\infty $ is compatible with
infinite energy resolution (since $\bar{\Delta}t\leqslant \Delta t$, $\Delta
E=0$ implies $\Delta t=\infty $). However, the energy resolution has to do
with the experimental setup, so there might be no universal way to include
it. One may have to use different formulas, depending on the experiment.

Instead of going through this, we can try and guess how $\Delta E$ may
affect the results. Generically, $\Delta E$ can appear more or less
everywhere, but in most places it redefines quantities that are already
present, so we can neglect it. The $\Delta E$ dependence cannot be ignored
if it affects the imaginary part of the denominator of the propagator, which
survives the free-field limit. An effect of this type is not surprising, if
we consider that $\Delta E$ is associated with the time resolution $\bar{%
\Delta}t$, which must be compared with the particle lifetime, which in turn
is the reciprocal of the width.

The energy resolution we focus on is the one around the peak. Specifically,
we define the \textquotedblleft distance\textquotedblright\ in energy\ from
the peak as the ratio 
\begin{equation}
\frac{|p^{2}-m_{\text{ph}}^{2}|}{2m_{\text{ph}}}.  \label{resolu}
\end{equation}%
Then the (Lorentz invariant) meaning of $\Delta E$ is that we cannot resolve
momenta $p$ that are closer than $\Delta E$ to the peak.

In light of the remarks just made, we assume that when $\Delta E$ is
different from zero the predictions coincide with the ones we have written
above once we make the replacement%
\begin{equation}
\tilde{\epsilon}\rightarrow \tilde{\epsilon}+2m_{\text{ph}}\Delta E,
\label{repla}
\end{equation}%
after which we can take $\tilde{\epsilon}$ to zero, since at that point it
is no longer necessary.

The form of the $\Delta E$ dependence in (\ref{repla}) is not crucial, as
long as the correction vanishes when $\Delta E$ tends to zero. For example,
variants such as%
\begin{equation}
\tilde{\epsilon}\rightarrow \tilde{\epsilon}+\gamma m_{\text{ph}}\Delta
E\left( \frac{\Delta E}{m_{\text{ph}}}\right) ^{\delta },  \label{repla2}
\end{equation}%
with $\gamma >0$, $\delta >-1$, do not change the conclusions we derive.
Indeed, (\ref{repla}) vs (\ref{repla2}) is just a redefinition of $\Delta E$
into a function of $\Delta E$, which only affects the relations between $%
\Delta t$, $\bar{\Delta}t$ and $\Delta E$.

Making the replacement in formulas (\ref{Ms}) and (\ref{Md}) and letting $%
\tilde{\epsilon}$ tend to zero, we obtain%
\begin{eqnarray}
\Omega _{\varphi \hspace{0.01in}\text{particle}} &\simeq &\frac{2m_{\text{ph}%
}Z\Delta E}{(p^{2}-m_{\text{ph}}^{2})^{2}+m_{\text{ph}}^{2}(2\Delta E+\Gamma
)^{2}},  \label{1} \\
\Omega _{\varphi \hspace{0.01in}\text{decay}} &\simeq &\frac{m_{\text{ph}%
}Z\Gamma }{(p^{2}-m_{\text{ph}}^{2})^{2}+m_{\text{ph}}^{2}(2\Delta E+\Gamma
)^{2}},  \label{2}
\end{eqnarray}%
The results show that $\Omega _{\varphi \hspace{0.01in}\text{particle}}$ is
no longer zero.

From the phenomenological side, we may distinguish three cases:

--- Case of the $Z$ boson. Here $\Delta E\ll \Gamma /2$, so 
\begin{equation*}
\Omega _{\varphi \hspace{0.01in}\text{particle}}\simeq 0,\qquad \Omega
_{\varphi \hspace{0.01in}\text{decay}}\simeq \frac{m_{\text{ph}}Z\Gamma }{%
(p^{2}-m_{\text{ph}}^{2})^{2}+m_{\text{ph}}^{2}\Gamma ^{2}}.
\end{equation*}%
In agreement with experiment, the predictions tell us that we do not see the
particle: we see its decay products. The results do not depend on $\Delta E$
to the first degree of approximation.

--- Case of the muon. Here $m_{\text{ph}}\gg \Delta E\gg \Gamma /2$, so%
\begin{equation}
\Omega _{\varphi \hspace{0.01in}\text{particle}}\simeq \frac{2m_{\text{ph}%
}Z\Delta E}{(p^{2}-m_{\text{ph}}^{2})^{2}+4m_{\text{ph}}^{2}\Delta E^{2}}%
\simeq \pi Z\delta (p^{2}-m_{\text{ph}}^{2}),\qquad \Omega _{\varphi \hspace{%
0.01in}\text{decay}}\simeq 0.  \label{mu}
\end{equation}%
These results do not depend on $\Delta E$ to the first degree of
approximation and tell us that we see the particle, not its decay products,
in agreement with experiment.

--- Intermediate situations. When $\Delta E$ and $\Gamma $ are comparable,
we see both the particle and its decay products. The results depend on $%
\Delta E$ and so does the ratio $\Omega _{\varphi \hspace{0.01in}\text{%
particle}}/\Omega _{\varphi \hspace{0.01in}\text{decay}}=2\Delta E/\Gamma .$

We have uncovered, among the other things, that it is impossible to explain
the observation of the muon, i.e., derive $\Omega _{\varphi \hspace{0.01in}%
\text{particle}}\simeq \pi Z\delta (p^{2}-m_{\text{ph}}^{2})$ in (\ref{mu}),
without introducing the resolution $\Delta E$, even if the final result is
independent of $\Delta E$ (because the approximations valid for the muon
make\ it disappear). This fact can be understood as a consequence of the
energy-time uncertainty relation $\Delta E\hspace{0.01in}\sim 1/\bar{\Delta}%
t $. Indeed, $\Delta E=0$ implies an infinite time uncertainty, during which
every unstable particle has enough time to decay before being observed: it
is impossible to observe an unstable particle with infinite resolving power
on its energy.

This impossibility can also be appreciated by studying the limit $\epsilon
\rightarrow 0$ term by term, before the resummation. In subsection \ref%
{physpeak} and in appendix \ref{split}, we show that each term of the
expansions of $\Omega _{\varphi \hspace{0.01in}\text{particle}}$ and $\Omega
_{\varphi \hspace{0.01in}\text{decay}}$ is ill defined at $\Delta E=0$,
apart from a few ones: only the expansion of the sum $\Omega _{\varphi 
\hspace{0.01in}\text{particle}}+\Omega _{\varphi \hspace{0.01in}\text{decay}%
} $ is regular. Such a complication, which may sound surprising at first, is
actually unavoidable, precisely because a regular $\Omega _{\varphi \hspace{%
0.01in}\text{particle}}$ at $\Delta E=0$ would violate the energy-time
uncertainty relation, in the case of unstable particles. Quantum field
theory knows it, so to speak, and retaliates by creating problems in the
crucial places. These properties emphasize that the issues we are discussing
here concern not only new or unusual quantization prescriptions, but also
physical particles, although in different ways.

Another important observation is that the replacement (\ref{repla}) violates
unitarity at nonvanishing $\Delta E$. Indeed, a propagator (\ref{propag})
with $\epsilon \rightarrow \epsilon +2m_{\text{ph}}Z^{-1}\Delta E$\ comes
from a non-Hermitian Lagrangian. When the experimental setup has an
important impact on the predictions, we cannot consider the system as an
isolated one, so unitarity as we normally understand it may not hold.
Nevertheless, as long as $\Delta E$ tends to zero after the expansion in
powers of $\Gamma $, the replacement (\ref{repla}) is precisely the right
one to have perturbative unitarity, because $\Delta E$ plays the role of the
infinitesimal width $\epsilon $.

\subsection{Conditions of convergence}

Let us discuss the convergence radius after the replacement (\ref{repla}).
Since $\Delta E$ is independent of the interactions (it affects the
propagator already at the tree level), the condition (\ref{converga}) for
convergence is 
\begin{equation}
\frac{(\Delta m^{2})^{2}+m_{\text{ph}}^{2}\Gamma ^{2}}{(p^{2}-m^{2})^{2}+4m_{%
\text{ph}}^{2}\Delta E^{2}}<1.  \label{radius}
\end{equation}%
It is possible to satisfy this condition for every $p$ if and only if%
\footnote{%
This is necessary if we want to use the dressed propagator inside bigger
diagrams, which involves integrals over $p$, or integrate on the whole phase
space of final states.}%
\begin{equation}
\Delta E>\frac{1}{2m_{\text{ph}}}\sqrt{m_{\text{ph}}^{2}\Gamma ^{2}+(\Delta
m^{2})^{2}}\equiv \Delta E_{\text{min}}.  \label{unce}
\end{equation}

If we assume that the powers of $\Delta m^{2}$ are resummed by default into
the physical mass $m_{\text{ph}}$, we can use (\ref{repla}) inside (\ref%
{converga2}), so the convergence condition becomes 
\begin{equation}
\frac{m_{\text{ph}}^{2}\Gamma ^{2}}{(p^{2}-m_{\text{ph}}^{2})^{2}+4m_{\text{%
ph}}^{2}\Delta E^{2}}<1,  \label{radius2}
\end{equation}%
which holds for every $p$ if and only if%
\begin{equation}
\Delta E>\frac{\Gamma }{2}\equiv \Delta E_{\text{min}}.  \label{unce2}
\end{equation}

The validity of the resummation can be extended to the peak region, i.e.,
beyond the bounds (\ref{unce}) or (\ref{unce2}), by means of analyticity.
Yet, it is crucial for the discussion that follows to remember that there is
a region where the convergence is guaranteed and a region that can only be
reached by means of analyticity. When analyticity does not hold, the peak
region cannot be reached.

Note that the left-hand sides of (\ref{radius}) and (\ref{radius2})\ are the
expansion parameters. When the inequalities (\ref{radius}) or (\ref{radius2}%
) do not hold, the problem is nonperturbative, not because the coupling is
large, but because the expansion parameter is large, due to its $p$
dependence. Specifically, if $\Delta E$ vanishes or is too small, (\ref%
{radius}) and (\ref{radius2}) are violated in the peak region $|p^{2}-\tilde{%
m}^{2}|\leqslant 2m_{\text{ph}}\Delta E_{\text{min}}$ no matter how small
the coupling is, where $\tilde{m}$ is $m$ or $m_{\text{ph}}$ depending on
the case.

\subsection{Ill-defined sums with physical particles}

\label{physpeak}

Now we show that ill-defined distributions also concern physical (unstable)
particles, when their observation is considered separately from the
observation of their decay products. To achieve this goal, we study the
limit $\epsilon \rightarrow 0$ term by term in the expansions of $\Omega
_{\varphi \hspace{0.01in}\text{particle}}$ and $\Omega _{\varphi \hspace{%
0.01in}\text{decay}}$, before their resummations.

The perturbative expansion of $\Omega _{\varphi \hspace{0.01in}\text{particle%
}}$ reads, from formula (\ref{omica1}), 
\begin{equation*}
\frac{\epsilon }{|p^{2}-m^{2}-\Sigma +i\epsilon |^{2}}=\frac{\epsilon }{%
(p^{2}-m^{2})^{2}+\epsilon ^{2}}+2\frac{\epsilon (p^{2}-m^{2})\text{Re}%
[\Sigma ]+\epsilon ^{2}\text{Im}[\Sigma ]}{((p^{2}-m^{2})^{2}+\epsilon
^{2})^{2}}+\mathcal{O(}\Sigma ^{2}).
\end{equation*}%
The zeroth order is $\pi \delta (p^{2}-m^{2})$, while the $\mathcal{O(}%
\Sigma )$ term proportional to Re$[\Sigma ]$ gives $\delta ^{\prime
}(p^{2}-m^{2})$. It is easy to check that the $\mathcal{O(}\Sigma )$ term
proportional to Im$[\Sigma ]$ is not a distribution. Something similar
happens at higher orders. Using the approximation (\ref{approx}) and
concentrating on the expansion in powers of $\Gamma $, we obtain, from (\ref%
{Ms}),%
\begin{equation}
\Omega _{\varphi \hspace{0.01in}\text{particle}}\simeq \pi Z\delta (p^{2}-m_{%
\text{ph}}^{2})-\frac{2\tilde{\epsilon}^{2}m_{\text{ph}}Z\Gamma }{((p^{2}-m_{%
\text{ph}}^{2})^{2}+\tilde{\epsilon}^{2})^{2}}+\mathcal{O}(\Gamma ^{2}).
\label{spart}
\end{equation}

The practical consequence is that, when we make the replacement (\ref{repla}%
) and set $\epsilon =0$, we get corrections that are very large for $\Delta
E\ll \Gamma /2$. The ill-defined distributions (or large contributions) just
described mutually cancel in the sums (\ref{1}) plus (\ref{2}), (\ref{Ms})
plus (\ref{Md}) and (\ref{omica1}) plus (\ref{omica2}).

The technical problems arise because we are trying to separate the
observation of the particle from the observation of its decay products.
Since a propagator is the sum of a Cauchy principal value and a Dirac delta
function, arbitrary powers of such distributions are generated. A way to
define them is by means of the symmetric coincidence-splitting method
described in appendix \ref{split}. Working out the two lines of fig. \ref%
{cutdiagrams} with this method, we obtain formulas (\ref{omica3}) for $%
\Omega _{\varphi \hspace{0.01in}\text{particle}}$ and $\Omega _{\varphi 
\hspace{0.01in}\text{decay}}$. The sum $\Omega _{\varphi \hspace{0.01in}%
\text{particle}}+\Omega _{\varphi \hspace{0.01in}\text{decay}}$ is well
defined and coincides with what we expect, as shown in equation (\ref{bw}),
but $\Omega _{\varphi \hspace{0.01in}\text{particle}}$ and $\Omega _{\varphi 
\hspace{0.01in}\text{decay}}$, taken separately, are ill-defined
distributions, similar to $\Delta _{\hat{\Gamma}}$.

Finally, formula (\ref{2}) gives%
\begin{equation*}
\Omega _{\varphi \hspace{0.01in}\text{decay}}\simeq \frac{\Gamma }{2\Delta E}%
\frac{2m_{\text{ph}}Z\Delta E}{(p^{2}-m_{\text{ph}}^{2})^{2}+4m_{\text{ph}%
}^{2}\Delta E^{2}}
\end{equation*}%
for $\Gamma \ll 2\Delta E$, which shows that we cannot observe the
\textquotedblleft peak of the muon\textquotedblright . Instead, we observe a
much smaller bump entirely due to the nonvanishing energy resolution of our
experimental setup.

\section{Dressed propagator of purely virtual particles}

\label{fakeons}\setcounter{equation}{0}

In this section we investigate the resummation of the self-energies in the
case of purely virtual particles $\chi $. The contribution to the amplitude
is $\mathcal{M}_{\chi }=i\hat{P}_{\chi }$ and $\text{Im}[\mathcal{M}_{\chi
}]=\text{Re}[\hat{P}_{\chi }]\simeq \Omega _{\chi \hspace{0.01in}\text{%
\textquotedblleft decay\textquotedblright }}$: there is no $\Omega _{\chi 
\hspace{0.01in}\text{\textquotedblleft particle\textquotedblright }}$ in the
case of fakeons. Indeed, the diagrams shown in the first line of fig. \ref%
{cutdiagrams} vanish identically, because a fakeon does not appear among the
final states (its cut propagator being zero). For the same reason, $\Omega
_{\chi \hspace{0.01in}\text{\textquotedblleft decay\textquotedblright }}$ is
not really associated with the decay of the fakeon and $\Gamma $ has a
different interpretation, which we provide below. Unless specified
differently, we assume that $\Gamma $ is strictly positive, since $\Gamma =0$
implies that $\Omega _{\chi \hspace{0.01in}\text{\textquotedblleft
decay\textquotedblright }}$ also vanishes.

We include the energy resolution $\Delta E$ into the dressed propagators by
means of the replacement%
\begin{equation}
\tilde{\epsilon}\rightarrow \tilde{\epsilon}+m_{\text{ph}}\Delta E,
\label{replas}
\end{equation}
which differs from (\ref{repla}) by a factor 2, for reasons that become
clear in a moment. After letting $\epsilon $ tend to zero, we obtain%
\begin{eqnarray}
\hat{P}_{\chi } &\simeq &\frac{iZ(p^{2}-m^{2})}{(p^{2}-m^{2})(p^{2}-m_{\text{%
ph}}^{2}+im_{\text{ph}}\Gamma )+m_{\text{ph}}^{2}\Delta E^{2}},  \notag \\
\hat{P}_{\chi } &\simeq &\frac{iZ(p^{2}-m_{\text{ph}}^{2})}{(p^{2}-m_{\text{%
ph}}^{2})(p^{2}-m_{\text{ph}}^{2}+im_{\text{ph}}\Gamma )+m_{\text{ph}%
}^{2}\Delta E^{2}}.  \label{Mfake}
\end{eqnarray}%
from (\ref{phatf}) and (\ref{phatf2}), respectively.

As before, the corrections due to $Z$ can be resummed straightforwardly. As
for the others, we use the first formula of (\ref{Mfake}) if we want to
treat $\Delta m^{2}$ and $\Gamma $ on an equal footing, and the second
formula of (\ref{Mfake}) if we assume that the powers of $\Delta m^{2}$ are
resummed by default into $m_{\text{ph}}^{2}$. The convergence conditions are
then%
\begin{equation}
\frac{\sqrt{m_{\text{ph}}^{2}\Gamma ^{2}+(\Delta m^{2})^{2}}\left\vert
p^{2}-m^{2}\right\vert }{(p^{2}-m^{2})^{2}+m_{\text{ph}}^{2}\Delta E^{2}}%
<1,\qquad \frac{m_{\text{ph}}\Gamma |p^{2}-m_{\text{ph}}^{2}|}{(p^{2}-m_{%
\text{ph}}^{2})^{2}+m_{\text{ph}}^{2}\Delta E^{2}}<1,  \label{conve}
\end{equation}%
respectively. In turn, (\ref{conve}) hold for every $p$ if and only if%
\begin{equation}
\Delta E>\frac{1}{2m_{\text{ph}}}\sqrt{m_{\text{ph}}^{2}\Gamma ^{2}+(\Delta
m^{2})^{2}}\equiv \Delta E_{\text{min}},\qquad \Delta E>\frac{\Gamma }{2}%
\equiv \Delta E_{\text{min}}.  \label{indet}
\end{equation}

If $\Delta E\leqslant \Delta E_{\text{min}}$ the resummation is justified
for momenta $p$ such that%
\begin{equation}
|p^{2}-\tilde{m}^{2}|>m_{\text{ph}}\Delta E_{\text{min}}\left( 1+\sqrt{%
1-R^{2}}\right) ,  \label{bonde2}
\end{equation}%
where $R=\Delta E/\Delta E_{\text{min}}$ and $\tilde{m}$ is $m$ or $m_{\text{%
ph}}$ depending on the case. Formally, it is also convergent for 
\begin{equation}
|p^{2}-\tilde{m}^{2}|<m_{\text{ph}}\Delta E_{\text{min}}\left( 1-\sqrt{%
1-R^{2}}\right) \leqslant m_{\text{ph}}\Delta E,  \label{bonde3}
\end{equation}%
but the size of this region is smaller than the size of the region we can
resolve, so we ignore it.

We cannot trust the resummation when (\ref{bonde2}) does not hold, which
means sufficiently close to the peak\footnote{%
Formulas (\ref{Mfake}) give a couple of bumps, rather than a peak, as shown
in fig. \ref{Fpeaks}. Also note that the first expression depends on both $%
m^{2}$ and $m_{\text{ph}}^{2}$. The points $p^{2}=m^{2}$ and $p^{2}=m_{\text{%
ph}}^{2}$ are within the range $\Delta E_{\text{min}}$. That said, we keep
referring to the region $p^{2}\simeq m^{2}$, $p^{2}\simeq m_{\text{ph}}^{2}$
by calling it \textquotedblleft peak region\textquotedblright .}. When $%
\Delta E$ tends to zero (infinite resolving power), we cannot trust it for $%
|p^{2}-\tilde{m}^{2}|\leqslant 2m_{\text{ph}}\Delta E_{\text{min}}$.

These arguments show that in the case of purely virtual particles the
distance in energy from the peak is defined by the ratio 
\begin{equation}
\frac{|p^{2}-\tilde{m}^{2}|}{2m_{\text{ph}}}.  \label{resolut}
\end{equation}%
They also justify the absence of the factor 2 in (\ref{replas}) with respect
to (\ref{repla}). As before, the Lorentz invariant definition of $\Delta E$
is the minimum distance we can resolve.

The denominator of $\hat{P}_{\chi }$ never vanishes, but the fakeon
prescription is not analytic, so we cannot advocate analyticity to cross the
boundary of the convergence region. This means that, strictly speaking, the
result of the resummation is infinite when (\ref{conve}) does not hold.
There, the problem is nonperturbative, since the $p$ dependence makes the
expansion parameter too large around $p^{2}\simeq \tilde{m}^{2}$.

We recall that in \cite{diagrammarMio} it was shown that the introduction of
the energy resolution $\Delta E$ for fakeons does not violate the optical
theorem.

\subsection{Peak uncertainty}

The nonperturbative theory, assuming that it exists, may remove the obstacle 
$\Delta E_{\text{min}}$ and return a unique answer when the inequality (\ref%
{indet}) is violated. It might also involve new mathematics and turn (\ref%
{indet}) into a matter of principle, like a new uncertainty relation. In
that case, its physical meaning would be that, no matter how precisely we
determine the energy of the initial states, the process will be insensitive
to any improvement beyond the limit (\ref{indet}) and return a plot similar
to the one we obtain with a lower resolution. We say more about this in the
section \ref{phenom}. See also section \ref{microc} for comments of the
(lack of) relation between the uncertainty (\ref{indet}) and the violation
of microcausality, typical of fakeons.

We have already remarked that the origin of the problems is a nonvanishing $%
\Gamma $, while $\Delta m^{2}$ plays a secondary role. For this reason, we
have a preference for the option (\ref{phatf2}) and the peak uncertainty 
\begin{equation}
\Delta E>\frac{\Gamma }{2}.  \label{indeto}
\end{equation}%
Since $\Gamma $ and $\Delta m^{2}$ are typically of the same order, the
bounds (\ref{indet}) and (\ref{indeto}) are generically equivalent.
Nevertheless, when $\Gamma $ is identically zero, as in the models of ref. 
\cite{Tallinn1}, (\ref{indeto}) predicts no uncertainty, with a dressed
propagator given by formula (\ref{phdressed}) everywhere. Instead, the left
formulas of (\ref{conve}) and (\ref{indet}) predict a peak uncertainty also
in that case. In the absence of knowledge about the nonperturbative sector
of the theory, only experiments can decide between the two possibilities.

A further comment concerns asymptotic series. We know that, normally,
perturbative quantum field theory provides predictions in the form of
asymptotic series. In most cases the first few terms of the series decrease
(in modulus), up to a certain order $n_{\text{asy}}$, and then start to
increase uncontrollably. If we truncate the expansion to $n_{\text{asy}}$ or
a lower order, we typically get precise predictions of the experimental
results. Sometimes, as for the muon anomalous magnetic moment, they are
impressively precise.

In the case of the fakeon self-energy, if we view the perturbative expansion
as an asymptotic series, we have to conclude that close enough to the peak\
we cannot keep any terms, not even the lowest nonvanishing order: when the
expansion parameter (\ref{conve}) is larger than one, the series is only
made of increasing terms, so\ no $n_{\text{asy}}$ exists. In section \ref%
{phenom} we argue what the missing part might look like phenomenologically.

\subsection{Fake ghosts}

Finally, we comment on the \textquotedblleft fake ghosts\textquotedblright ,
which are the fakeons obtained by changing the quantization prescription of
ghosts into the fakeon one. The tree-level propagator of a leg that, if
broken, disconnects the diagram is%
\begin{equation*}
P_{\chi }=-\frac{i(p^{2}-m^{2})}{(p^{2}-m^{2})^{2}+\epsilon ^{2}}.
\end{equation*}%
It is convenient to use the approximation (\ref{approxgh}) around $%
p^{2}\simeq m^{2}$. Note that we still have $\Gamma \geqslant 0$, since the
fake ghosts obey the optical theorem, like any other type of fakeons. The
resummation gives%
\begin{equation*}
\hat{P}_{\chi }\simeq -\frac{iZ(p^{2}-m^{2})}{(p^{2}-m^{2})(p^{2}-m_{\text{ph%
}}^{2}-im_{\text{ph}}\Gamma )+\tilde{\epsilon}^{2}},\quad \hat{P}_{\chi
}\simeq -\frac{iZ(p^{2}-m_{\text{ph}}^{2})}{(p^{2}-m_{\text{ph}%
}^{2})(p^{2}-m_{\text{ph}}^{2}-im_{\text{ph}}\Gamma )+\tilde{\epsilon}^{2}},
\end{equation*}%
in the two cases treated in subsection \ref{purely}, where $\tilde{\epsilon}%
=\epsilon Z^{1/2}$. We see that $P_{\chi }$ and $\hat{P}_{\chi }$ are just
the complex conjugates of the ones we had before. Therefore, the properties
of one type of fakeons are also valid for the other type. In the next
section, we continue with the fakeons obtained by changing the quantization
prescription of physical particles.

\section{Dressed propagator of ghosts}

\label{ghosts}\setcounter{equation}{0}

In this section we investigate the resummation of self-energies in the case
of ghosts $\phi $. The amplitude $\mathcal{M}_{\phi }=i\hat{P}_{\phi }$
gives Im$[\mathcal{M}_{\phi }]\simeq \Omega _{\phi \hspace{0.01in}\text{ghost%
}}+\Omega _{\phi \hspace{0.01in}\text{decay}}$, close to the peak, with%
\begin{equation*}
\Omega _{\phi \hspace{0.01in}\text{ghost}}\simeq -\frac{\tilde{\epsilon}Z}{%
(p^{2}-m_{\text{ph}}^{2})^{2}+(\tilde{\epsilon}-m_{\text{ph}}\Gamma )^{2}}%
,\qquad \Omega _{\phi \hspace{0.01in}\text{decay}}\simeq \frac{m_{\text{ph}%
}\Gamma Z}{(p^{2}-m_{\text{ph}}^{2})^{2}+(\tilde{\epsilon}-m_{\text{ph}%
}\Gamma )^{2}}.
\end{equation*}%
When $\epsilon $ tends to zero, we obtain%
\begin{equation}
\Omega _{\phi \hspace{0.01in}\text{ghost}}\simeq 0,\qquad \Omega _{\phi 
\hspace{0.01in}\text{decay}}\simeq \frac{m_{\text{ph}}Z\Gamma }{(p^{2}-m_{%
\text{ph}}^{2})^{2}+m_{\text{ph}}^{2}\Gamma ^{2}}.  \label{wrong0}
\end{equation}%
and the classical limit $\Gamma \rightarrow 0$, $Z\rightarrow 1$, gives%
\begin{equation}
\Omega _{\phi \hspace{0.01in}\text{ghost}}\simeq 0,\qquad \Omega _{\phi 
\hspace{0.01in}\text{decay}}\simeq \pi \delta (p^{2}-m_{\text{ph}}^{2}).
\label{wrong}
\end{equation}%
Formulas (\ref{wrong0}) and (\ref{wrong}) show that $\Omega _{\phi \hspace{%
0.01in}\text{ghost}}$ and $\Omega _{\phi \hspace{0.01in}\text{decay}}$
coincide with the $\Omega _{\varphi \hspace{0.01in}\text{ghost}}$ and $%
\Omega _{\varphi \hspace{0.01in}\text{decay}}$ we obtain from a physical
particle, given in formula (\ref{limpa}). Yet, the classical limit should
return what we started from (i.e., a ghost, not a physical particle).

The solution of this puzzle mimics the solution of the muon problem given in
subsection \ref{muonprobl}. Inserting the energy resolution $\Delta E$ by
means of the replacement (\ref{repla}), we obtain%
\begin{equation}
\hat{P}_{\phi }\simeq -\frac{iZ}{p^{2}-m_{\text{ph}}^{2}+i(\tilde{\epsilon}%
+2m_{\text{ph}}\Delta E-m_{\text{ph}}\Gamma )},  \label{pghost}
\end{equation}%
so the limit $\epsilon \rightarrow 0$ gives%
\begin{equation*}
\Omega _{\phi \hspace{0.01in}\text{ghost}}\simeq -\frac{2m_{\text{ph}%
}Z\Delta E}{(p^{2}-m_{\text{ph}}^{2})^{2}+m_{\text{ph}}^{2}(2\Delta E-\Gamma
)^{2}},\qquad \Omega _{\phi \hspace{0.01in}\text{decay}}\simeq \frac{m_{%
\text{ph}}Z\Gamma }{(p^{2}-m_{\text{ph}}^{2})^{2}+m_{\text{ph}}^{2}(2\Delta
E-\Gamma )^{2}}.
\end{equation*}

In the classical limit, $\Gamma $ tends to zero, but $\Delta E$ remains
constant. In the regime $m_{\text{ph}}\gg \Delta E\gg \Gamma /2$, we get%
\begin{equation}
\Omega _{\phi \hspace{0.01in}\text{ghost}}\simeq -\pi Z\delta (p^{2}-m_{%
\text{ph}}^{2}),\qquad \Omega _{\phi \hspace{0.01in}\text{decay}}\simeq 0.
\label{mughost}
\end{equation}%
The result, which is independent of $\Delta E$, is now correct, since, as
expected, it gives back the ghost we started from.

Again, we have been able to obtain the correct classical limit (\ref{mughost}%
) only by inserting the resolution $\Delta E$ and removing it afterward.
Before introducing $\Delta E$, we are lead to think that we can convert a
ghost into a sort of physical particle by turning on interactions, resumming
them and letting $\hbar $ tend to zero at the end. As soon as we introduce $%
\Delta E$, we see that this is not possible.

To better identify the obstruction that makes it impossible, we analyze the
validity of the resummation. From 
\begin{equation*}
\hat{P}_{\phi }\simeq -\frac{iZ}{p^{2}-m^{2}+2im_{\text{ph}}\Delta E}\frac{1%
}{1-a},\qquad a=\frac{im_{\text{ph}}\Gamma +\Delta m^{2}}{p^{2}-m^{2}+2im_{%
\text{ph}}\Delta E},
\end{equation*}%
we see that the convergence radius $|a|<1$ gives back the condition (\ref%
{radius}) we found for physical particles, which is satisfied for every $p$
if and only if (\ref{unce}) holds. If we assume that the powers of $\Delta
m^{2}$ are resummed by default into $m_{\text{ph}}^{2}$, we get the
convergence condition (\ref{radius2}), which holds for every $p$ if and only
if (\ref{unce2}) holds.

If we want a better resolution, we can try and reach a $\Delta E<\Delta E_{%
\text{min}}$ starting from values that satisfy $\Delta E>\Delta E_{\text{min}%
}$. When we lower $\Delta E$, everything is fine as long as $\Delta E$
remains larger than $\Gamma/2 $, but if we want to reduce $\Delta E$ even
more, we end by crossing the boundary $\Delta E=\Gamma /2$, where the
dressed propagator (\ref{pghost}) is not well prescribed. Indeed, at $\tilde{%
\epsilon}=0$ it reads%
\begin{equation*}
\hat{P}_{\phi }\simeq -\frac{i}{p^{2}-m_{\text{ph}}^{2}}.
\end{equation*}%
A nonvanishing $i\tilde{\epsilon}$ is not of help here, since it just shifts
the boundary of the region. Thus, analyticity does not allow us to reach the
domain $\Delta E<\Gamma /2$.

We can in principle stretch the argument as follows. The denominator of $%
\hat{P}_{\phi }$ does not vanish at $\Delta E=\Gamma /2$ away from the peak.
Thus, as long as we stay away from the peak, we can get as close as we wish
to it by analyticity.\ If we exclude an arbitrarily small neighborhood of
the peak, we should be able to use the resummed formula (\ref{resugh}) (with 
$\tilde{\epsilon}=0$).

Nevertheless, the very fact that we cannot include the peak tells us that we
are missing something there. If what we are missing is a finite sum of
contact terms ($\delta $ functions and derivatives of $\delta $ functions),
we can get rid of them by excluding an arbitrarily small neighborhood of the
peak. In that case, formula (\ref{resugh}) works well.

However, in section \ref{differences} we showed that what is missing is an
infinite sum of contact terms. Then, their effects can extend around the
peak. For example, any regular function $f(x)$ that decreases exponentially
at infinity can be written as an infinite sum of contact terms\footnote{%
The formula follows from the Taylor expansion of $\delta (x-y)$ at $x$
inside $f(x)=\int \mathrm{d}y\hspace{0.01in}f(y)\delta (x-y)$.
Alternatively, it is enough to require that the Fourier transform $\tilde{f}%
(p)$ of $f(x)$ can be expanded as a power series, $\tilde{f}%
(p)=\sum_{n=0}^{\infty }f_{n}(ip)^{n}$. The simplest example is the Gaussian
function $\tilde{f}(p)=\exp (-p^{2}/2)$.}: 
\begin{equation*}
f(x)=\sum_{n=0}^{\infty }f_{n}\delta ^{(n)}(x),\qquad f_{n}=\frac{1}{n!}%
\int_{-\infty }^{+\infty }\mathrm{d}y\hspace{0.01in}f(y)(-y)^{n}.
\end{equation*}

\section{Phenomenology of fake particles}

\label{phenom}\setcounter{equation}{0}

In this section, we discuss the phenomenological aspects of the results
obtained on section \ref{fakeons} about purely virtual particles. The two
regimes studied there, the one of the $Z$ boson and the one of the muon, are
mirrored into the ones of collider physics and quantum gravity,
respectively. We also investigate effective dressed propagators to describe
the nonperturbative contributions that are activated in the peak\ region.

We begin by emphasizing that there always exist physical configurations
where the nonperturbative effects are avoided. It is sufficient to restrict
the invariant masses $M=\sqrt{p^{2}}$ of the subsets of external states
mediated by a fakeon so as to stay away from the region of the fakeon peak.
Because of (\ref{bonde2}), if such invariant masses satisfy%
\begin{equation}
|M^{2}-\tilde{m}^{2}|>2m_{\text{ph}}\Delta E_{\text{min}},  \label{conda}
\end{equation}%
where $\tilde{m}$ is $m$ or $m_{\text{ph}}$, depending on the case, we can
take $\Delta E$ to zero in formulas (\ref{Mfake}). Under assumptions like (%
\ref{conda}), we can make predictions about scattering processes at
arbitrarily high energies.

\begin{figure}[t]
\begin{center}
\includegraphics[width=6truecm]{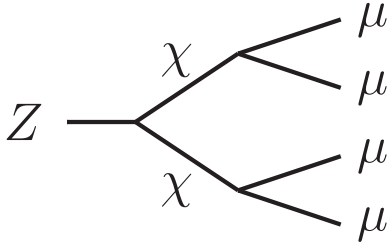}
\end{center}
\caption{$Z$ decay into four muons mediated by two fakeons $\protect\chi $}
\label{Z4mu}
\end{figure}

For example, in a hypothetical decay $Z\rightarrow 4\mu $ mediated by two
neutral fakeons $\chi $ of mass $m$, shown in fig. \ref{Z4mu}, the
conditions (\ref{conda}) must hold for all the pairs $\mu ^{+}\mu ^{-}$.
This process was considered in ref. \cite{Tallinn2}.

Conditions like (\ref{conda}), however, do not allow us to sum over the
whole phase spaces of the final states, because they include contributions
from the regions of the fakeon peaks. At the same time, a nonvanishing
resolution $\Delta E$, by conditions (\ref{indet}), is bad around the peaks.
The processes involving configurations beyond the bounds (\ref{conda}) and (%
\ref{indet}) can be investigated phenomenologically, starting from the
general properties of fakeons.

\subsection{Quantum gravity}

Quantum gravity is unitary and renormalizable if a few higher-derivative
terms are included (besides the Hilbert-Einstein term and the cosmological
term) and the extra degrees of freedom, which are a scalar field $\phi _{%
\text{infl}}$ and a spin-2 field $\chi _{\mu \nu }$, are quantized as a
physical particle (the inflaton) and a spin-2 purely virtual particle,
respectively \cite{LWgrav}. Primordial cosmology fixes the inflaton mass to $%
m_{\phi }\simeq 3\cdot 10^{13}$GeV through the spectrum of the scalar
fluctuations. The consistency of the fakeon prescription on a nontrivial
background gives the ABP bound $m_{\chi }>m_{\phi }/4$ on the fakeon mass $%
m_{\chi }$ and a sharp prediction on the tensor-to-scalar ratio $r$ ($%
0.4\lesssim 1000r\lesssim 3$) of the primordial fluctuations \cite{ABP}.

We want to show that the case of $\chi _{\mu \nu }$ is similar to the case
of the muon, rather than the one of the $Z$ boson. If we assume that $%
m_{\chi }$ and $m_{\phi }$ are of the same order, at energies comparable
with such masses we have \cite{Absograv} 
\begin{equation*}
E\simeq m_{\chi }\simeq 3\cdot 10^{13}\text{GeV},\qquad \Gamma _{\chi
}\simeq \frac{m_{\chi }^{3}}{M_{\text{Pl}}^{2}}\frac{N_{s}+6N_{f}+12N_{v}}{%
120}\simeq 4\cdot 10^{2}\text{GeV},
\end{equation*}%
where $N_{s}$ is the number of real scalar fields, $N_{f}$ is the number of
Dirac fermions plus one half the number of Weyl fermions and $N_{v}$ is the
number of gauge bosons of the standard model coupled to quantum gravity.

The key values concerning the $Z$ boson are $\Gamma _{Z}\simeq 2$GeV, $%
E\simeq m_{Z}\simeq 91$GeV, $\Delta E_{Z}\simeq $ $2\cdot 10^{-3}$GeV (the
experimental error on $m_{Z}$). Taking $\left\vert \Delta
m_{Z}^{2}\right\vert \simeq m_{Z}\Gamma _{Z}$, because they are of the same
order, we find $\Delta E_{Z\hspace{0.01in}\text{min}}/\Delta E_{Z}\simeq
10^{3}$, which violates (\ref{unce}), (\ref{unce2}). Thus, when we need to
treat the $Z$ boson around the peak, we must use its dressed propagator. We
know that we can do it, because the resummation of the self-energies in the
peak region is justified by analyticity.

Now, the fakeon width $\Gamma _{\chi }$ is $2\cdot 10^{2}$ times larger than
the $Z$ width $\Gamma _{Z}$. However, the $\chi _{\mu \nu }$ mass $m_{\chi }$
is $3\cdot 10^{11}$ times larger than $m_{Z}$. Using the $Z$ data as
reference values, we can assume that a hypothetical scattering process
involving $\chi _{\mu \nu }$ particles at energies $E\simeq m_{\chi }$ will
have an error $\Delta E_{\chi }\simeq $ $3\cdot 10^{11}\Delta E_{Z}$. Then
we expect%
\begin{equation}
\frac{\Delta E_{\chi \hspace{0.01in}\text{min}}}{\Delta E_{\chi }}\simeq
10^{-5},  \label{corregrav}
\end{equation}%
which fulfills (\ref{unce})-(\ref{unce2}). This is a situation where we do
not need to resum the series into (\ref{Mfake}): we can just use the
tree-level propagator (\ref{pQtree}) for $\chi _{\mu \nu }$.

As we have anticipated, the situation is similar to the one of the muon,
since $m_{\chi }\gg \Delta E_{\chi }\gg \Gamma _{\chi }/2,\Delta E_{\chi 
\hspace{0.01in}\text{min}}$. The result depends on $\Delta E_{\chi }$ very
little. Precisely, the first propagator of (\ref{Mfake}) around $p^{2}-m^{2}$
($m^{2}$ being now $m_{\chi }^{2}$) is%
\begin{eqnarray}
&&\frac{iZ_{\chi }(p^{2}-m_{\chi }^{2})}{(p^{2}-m_{\chi }^{2})^{2}+m_{\text{%
ph}}^{2}\Delta E_{\chi }^{2}}\left[ 1+\frac{(p^{2}-m_{\chi }^{2})(\Delta
m_{\chi }^{2}-im_{\text{ph}}\Gamma _{\chi })}{(p^{2}-m_{\chi }^{2})^{2}+m_{%
\text{ph}}^{2}\Delta E_{\chi }^{2}}+\cdots \right]  \notag \\
&\sim &\mathcal{P}\frac{iZ_{\chi }}{p^{2}-m_{\chi }^{2}}+iZ_{\chi }\pi
\delta (p^{2}-m_{\chi }^{2})\frac{(\Delta m_{\chi }^{2}-im_{\text{ph}}\Gamma
_{\chi })}{2m_{\text{ph}}\Delta E_{\chi }}+\cdots ,  \label{correg}
\end{eqnarray}%
having used 
\begin{equation}
\lim_{\epsilon \rightarrow 0}\frac{\epsilon x^{2}}{(x^{2}+\epsilon ^{2})^{2}}%
=\lim_{\epsilon \rightarrow 0}\frac{\epsilon ^{3}}{(x^{2}+\epsilon ^{2})^{2}}%
=\frac{\pi }{2}\delta (x).  \label{distr}
\end{equation}%
Unless we can measure corrections such as (\ref{corregrav}), we can take $%
\Delta m_{\chi }^{2}$ and $\Gamma _{\chi }$ to $0$, which gives a propagator
equal to $i$ times the principal value of $Z_{\chi }/(p^{2}-m_{\chi }^{2})$.

For a better comparison, let us analyze the muon itself, where%
\begin{equation*}
\Gamma _{\mu }\sim 10^{-19}\text{GeV},\qquad \Delta E_{\mu }\sim 10^{-9}%
\text{GeV},\qquad \frac{\Gamma _{\mu }}{2\Delta E_{\mu }}\sim 10^{-10},
\end{equation*}%
$\Delta E_{\mu }$ being the error on the muon mass. Using formula (\ref%
{spart})\ with $\tilde{\epsilon}\rightarrow 2m_{\text{ph}}\Delta E$,
together with (\ref{distr}), we find%
\begin{equation*}
\Omega _{\mu \hspace{0.01in}\text{particle}}\simeq \pi Z\delta (p^{2}-m_{%
\text{ph}}^{2})\left( 1-\frac{\Gamma _{\mu }}{2\Delta E_{\mu }}\right)
+\cdots .
\end{equation*}%
Unless we can detect corrections $\sim 10^{-10}$, we can ignore the muon
width altogether in its propagator, as is normally done.

If some day we will be able to study scattering processes in quantum gravity
at energies around $10^{13}$GeV with enough precision to test corrections as
small as (\ref{corregrav}), we will see effects such as those of formula (%
\ref{correg}).

\subsection{Collider physics}

If lighter fakeons exist in nature and have an observable impact in collider
physics, a phenomenological description of the nonperturbative effects
activated in the peak region may be useful.

We concentrate on the first formula of (\ref{Mfake}), since the second one
can be seen as a particular case of it with $\Delta m^{2}=0$, $\tilde{m}%
=m=m_{\text{ph}}$. Consider the resummed propagators (\ref{Mfake}), with the
first condition (\ref{indet}). We cannot take $\Delta E$ to zero there,
otherwise (\ref{indet}) is violated. The simplest possibility is to consider
the \textquotedblleft minimum uncertainty propagator\textquotedblright ,
which is the one with $\Delta E=\Delta E_{\text{min}}$. Is this an
acceptable, $\Delta E$-independent phenomenological fakeon propagator? The
answer is no, because its classical limit $\Delta m^{2},\Gamma \rightarrow 0$
is proportional to%
\begin{equation}
\lim_{\varepsilon \rightarrow 0}\frac{ix}{x^{2}+2i\varepsilon \mathrm{e}%
^{i\theta }x+\varepsilon ^{2}}=\mathcal{P}\frac{i}{x}+\frac{\pi \mathrm{e}%
^{i\theta _{0}}}{\sqrt{1+\mathrm{e}^{2i\theta _{0}}}}\delta (x),\qquad x=%
\frac{p^{2}-m^{2}}{m_{\text{ph}}^{2}},  \label{result}
\end{equation}%
where $\theta =\arcsin (\Delta m^{2}/(2m_{\text{ph}}\Delta E_{\text{min}}))$%
, $\varepsilon =\Delta E_{\text{min}}/m_{\text{ph}}$ and $\theta _{0}$ is
the classical limit of $\theta $. Since $\Gamma >0$, the angle $\theta $
satisfies $|\theta |<\pi /2$. Moreover, assuming that $\Delta m^{2}$ and $%
\Gamma $ are of the same order, we also have $|\theta _{0}|<\pi /2$. The
limit (\ref{result}) can be proved by studying the even and odd parts
separately and acting on arbitrary test functions.

The result is not purely virtual, because the term proportional to $\delta
(x)$ signals a nontrivial on-shell contribution. The reason why the
coefficient of $\delta (x)$ is nonvanishing is that the classical limit is
incompatible with the equality $\Delta E=\Delta E_{\text{min}}$, because $%
\Delta E_{\text{min}}$ tends to zero, but the resolution $\Delta E$ does not
change in that limit.

The correct classical limit amounts to take $\Delta m^{2},\Gamma \rightarrow
0$ first and \textit{then} $\Delta E\rightarrow 0$, which indeed gives the
principal value. Moreover, by (\ref{indet}) $\Delta E$ should be strictly
larger than $\Delta E_{\text{min}}$, because $\Delta E=\Delta E_{\text{min}}$
does not belong to the convergence region.

Having learned that $\Delta E=\Delta E_{\text{min}}$ is not a good
phenomenological assumption, a better proposal that complies with the
requirements just outlined is%
\begin{equation}
\Delta E=\gamma \Delta E_{\text{min}}\left( \frac{m_{\text{ph}}}{\Delta E_{%
\text{min}}}\right) ^{\delta },\qquad \gamma >0,\qquad 0<\delta <1.
\label{putative}
\end{equation}%
The constants $\gamma $ and $\delta $ may be thought as the remnants of the
interaction with the experimental setup, or due to nonperturbative effects.
Then from (\ref{Mfake}) we have the effective propagator%
\begin{equation}
\hat{P}_{\chi }\simeq \frac{iZ(p^{2}-m^{2})}{(p^{2}-m^{2})(p^{2}-m_{\text{ph}%
}^{2}+im_{\text{ph}}\Gamma )+\gamma ^{2}m_{\text{ph}}^{2+2\delta }\Delta E_{%
\text{min}}^{2-2\delta }},  \label{dressedf}
\end{equation}%
which has the correct classical limit, since 
\begin{equation*}
\lim_{\varepsilon \rightarrow 0}\frac{ix}{x^{2}+2i\varepsilon \mathrm{e}%
^{i\theta }x+\gamma ^{2}\varepsilon ^{2-2\delta }}=\mathcal{P}\frac{i}{x}.
\end{equation*}%
This equality can be proved as before.

Formula (\ref{dressedf}) behaves correctly away from the peak, where it
approximates the Breit-Wigner formula%
\begin{equation}
\hat{P}_{\chi }\simeq \frac{iZ}{p^{2}-m_{\text{ph}}^{2}-im_{\text{ph}}\Gamma 
}.  \label{BWchi}
\end{equation}

\begin{figure}[t]
\begin{center}
\includegraphics[width=8truecm]{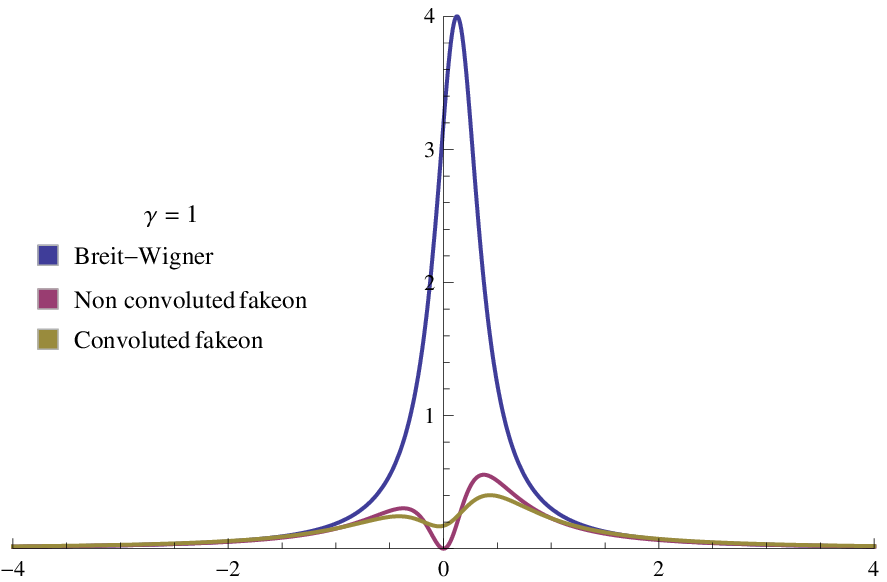} %
\includegraphics[width=8truecm]{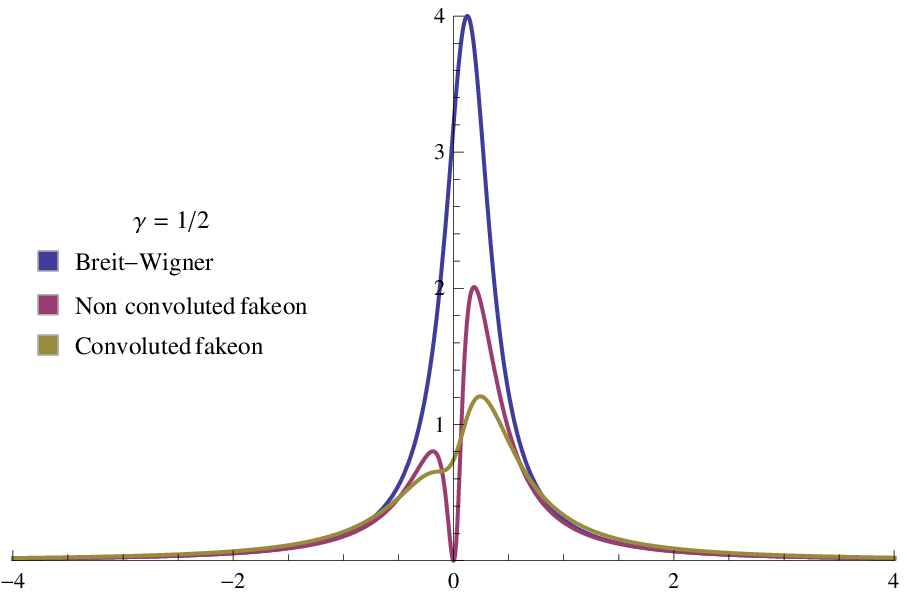}
\end{center}
\caption{Non-convoluted and convoluted fakeon plots with $\protect\gamma =1$
and $\protect\gamma =1/2$, compared to the Breit-Wigner plot of a standard
particle with the same width $\Gamma $ and the same $\Delta m^{2}$}
\label{Fpeaks}
\end{figure}

Presumably, the medium value $\delta =1/2$ is sufficiently accurate. The
plot of the real part of (\ref{dressedf}) is shown in fig. \ref{Fpeaks} (for 
$\delta =1/2$, $Z=1$, $m_{\text{ph}}=1$, $\Gamma =1/4$, $\Delta m^{2}=1/8$,
and for two choices of $\gamma $: $\gamma =1$ and $\gamma =1/2$). For
comparison, we include the plot of a Breit-Wigner peak for a standard
particle with the same width $\Gamma $ and the same $\Delta m^{2}$. The
fakeon plot better \textquotedblleft fills\textquotedblright\ the
Breit-Wigner one for lower values of $\gamma $. At $\Delta m^{2}=0$ the two
fakeon bumps are symmetric with respect to the vertical line. 

If $\Delta E$ is interpreted as an uncertainty or an experimental error, it
may be more accurate to consider the convolution%
\begin{equation*}
\tilde{P}_{\chi }(x)=\frac{m_{\text{ph}}\Delta E}{\pi }\int_{-\infty
}^{+\infty }\mathrm{d}y\frac{\hat{P}_{\chi }(x-y)}{y^{2}+m_{\text{ph}%
}^{2}\Delta E^{2}},
\end{equation*}%
where the expression (\ref{dressedf}) is seen as a function $\hat{P}_{\chi
}(x)$ of $x$. The convoluted dressed propagator is also shown in fig. \ref%
{Fpeaks}.

We see that, in general, the fakeon plot\ is suppressed with respect to the
one of a standard particle. This is not surprising, given the nature of a
fake particle. The three plots superpose as soon as we move away from the
peak region.

More generally, we can have an $x$-dependent factor $\gamma $ in formula (%
\ref{putative}), such as%
\begin{equation}
\gamma (x^{2})=\gamma _{0}\exp \left( -\sigma \frac{x^{2}m_{\text{ph}}}{%
\Delta E_{\text{min}}}\right) ,  \label{gammax}
\end{equation}%
where $\gamma _{0}$ and $\sigma $ are positive constants, so that $\Delta E$
practically vanishes away from the peak and the Breit-Wigner expression (\ref%
{BWchi}) is reached more rapidly. An expression like (\ref{gammax}) contains
essential singularities in the couplings and could be originated by
nonperturbative effects. With such a $\gamma $ we can extend (\ref{putative}%
) to $\delta =0$.

\bigskip

Following the line of thinking that leads to (\ref{dressedf}), we can search
for phenomenological\ formulas for ghosts as well. If we take, for example,%
\begin{equation*}
\Delta E=\gamma (x^{2})\frac{\Gamma }{2},\qquad \gamma (x^{2})>0,\text{\quad 
}\gamma (0)>1,\quad \gamma (\infty )=0,
\end{equation*}%
where $\gamma (x^{2})$ can be of the form (\ref{gammax}), we find, from (\ref%
{pghost}),%
\begin{equation}
\hat{P}_{\phi }\simeq -\frac{iZ}{p^{2}-m_{\text{ph}}^{2}+im_{\text{ph}%
}\Gamma (\gamma (x^{2})-1)},  \label{pgh}
\end{equation}%
which has the correct classical limit, since%
\begin{equation*}
-\lim_{\varepsilon \rightarrow 0}\frac{i}{x+i\varepsilon (\gamma (x^{2})-1)}%
=-i\mathcal{P}\frac{1}{x}-\pi \delta (x).
\end{equation*}%
Formula (\ref{pgh}) also behaves correctly away from the peak, where $\gamma
(x^{2})$ is negligible.

\section{Comparison between the peak uncertainty and the violation of
microcausality}

\label{microc}\setcounter{equation}{0}

Fakeons are responsible for the violation of microcausality \cite%
{Absograv,causalityQG}, which prevents predictions for time intervals
shorter than $\tau \equiv 1/m_{\chi }$. This means that the theory can be
tested only \textit{a posteriori}, after a delay $\tau $. In this section we
compare the violation of microcausality to the peak uncertainty of formula (%
\ref{indet}), which codifies the impossibility to get too close to the
fakeon peak. We find that, although the two have a common origin, they are
essentially different.

The violation of microcausality is associated with the intrinsic nonlocal
nature of the fakeon projection. Its effects can be appreciated in
coordinate space. We can illustrate them with an example taken from ref. 
\cite{causalityQG}. Consider the Lagrangian%
\begin{equation*}
\mathcal{L}(x,Q,t)=\frac{m}{2}\dot{x}^{2}-m\dot{x}\dot{Q}+\frac{m}{2\tau ^{2}%
}Q^{2}+xF_{\text{ext}}(t),
\end{equation*}%
where $x$ is the coordinate of a physical particle, $Q$ is the one of a
purely virtual particle, $m$ and $\tau $ are constants and $F_{\text{ext}%
}(t) $ is a time-dependent external force. The equations of motion give 
\begin{equation*}
\ddot{x}=-\frac{Q}{\tau ^{2}},\qquad m\ddot{Q}+\frac{m}{\tau ^{2}}Q=-F_{%
\text{ext}}(t).
\end{equation*}%
Since $Q$ is a fakeon, its equation admits the unique solution%
\begin{equation*}
mQ=-\mathcal{P}\frac{\tau ^{2}}{1+\tau ^{2}\frac{\mathrm{d}^{2}}{\mathrm{d}%
t^{2}}}F_{\text{ext}}(t)=-\frac{\tau }{2}\int_{-\infty }^{\infty }\mathrm{d}u%
\hspace{0in}\hspace{0.01in}F_{\text{ext}}(t-u)\sin \left( \frac{|u|}{\tau }%
\right) ,
\end{equation*}%
given by the fakeon prescription. Inserting this expression into the
equation of $x$, we obtain%
\begin{equation}
m\ddot{x}=\frac{1}{2\tau }\int_{-\infty }^{\infty }\mathrm{d}u\hspace{0in}%
\hspace{0.01in}F_{\text{ext}}(t-u)\sin \left( \frac{|u|}{\tau }\right) .
\label{equation}
\end{equation}%
The integral appearing on the right-hand side receives contributions from
the external force in the past as well as in the future. The amount of
future effectively contributing is 
\begin{equation}
\left\vert \Delta u\right\vert \simeq \tau ,  \label{uncecaus}
\end{equation}%
due to the oscillating behavior of $1/(2\tau )\sin (|u|/\tau )$, and
disappears for $\tau \rightarrow 0$, since $\lim_{\tau \rightarrow
0}1/(2\tau )\sin (|u|/\tau )=\delta (u)$. Thus, (\ref{uncecaus}) encodes the
fuzziness due to the violation of microcausality. It implies that we cannot
make predictions for time intervals shorter than $\tau $. However, we can,
in principle, check (\ref{equation}) a posteriori, if we manage to measure $%
x(t)$ and $F_{\text{ext}}(t)$ independently.

This example shows that the violation of microcausality due to fakeons does
not need a nonvanishing width. The key quantity encoding it is the fakeon
mass $m_{\chi }$ (which is $1/\tau $ in the toy model just considered). For
this reason, the violation survives the classical limit.

The peak uncertainty, instead, is encoded in the radiative corrections $%
\Gamma _{\chi }$ and possibly $\Delta m_{\chi }^{2}$, so it disappears in
the classical limit. It concerns the energy and implies that we cannot get
too close to $E\sim m_{\chi }$ in the channels mediated by fakeons. It does
not prevent predictions on processes occurring at higher energies. If light
fakeons exist in nature, it should be possible to detect the peak
uncertainty experimentally. Instead of seeing a resonance, as we expect for
a normal particle, we should see one or two bumps, with shapes that depend
on the experimental setup in a way that may be difficult, or impossible, to
predict.

While the violation of microcausality is always present, being associated
with the fakeon mass, it is possible to have no peak uncertainty (\ref%
{indeto}). It happens when the dressed propagator is the second of (\ref%
{Mfake}) and the fakeon width $\Gamma _{\chi }$ vanishes, as in the models
of ref. \cite{Tallinn1}.

These arguments show that there is no direct correspondence between the peak
uncertainty (\ref{indeto}) and the violation of microcausality.

\section{Conclusions}

\label{conclusions}\setcounter{equation}{0}

We have studied the dressed propagators of purely virtual particles and
compared the results with those of physical particles and ghosts, pointing
out several nontrivial issues that are commonly ignored. For example, the
usual dressed propagator is unable to explain the experimental observation
of long-lived unstable particles, like the muon. The difficulty can be
overcome by introducing the energy resolution $\Delta E$ of the experimental
setup. The need of $\Delta E$ is explained by the usual energy-time
uncertainty relation, which implies that it is impossible to observe an
unstable particle with infinite resolving power on the energy.

The expansions of the dressed propagators of physical particles, fake
particles and ghosts differ by infinite series of contact terms, which
cannot be summed into well-defined mathematical distributions. If we insist
in trusting the formal resummations, we obtain physical absurdities on new
types. Problematic sums also appear with physical particles, when we want to
distinguish their observation from the observation of their decay products.

The problems originate from the resummation of the self-energies in the peak
region, which lies outside the convergence domain of the geometric series.
In the case of physical particles, analyticity allows us to extend the
dressed propagator from the convergence region to the entire domain of
external momenta, provided we treat the observation of the particle and the
observation of its decay products as a whole. Analyticity is helpless in the
other cases. In particular, it is helpless in the cases of fake particles
and ghosts, where the formal sums cannot be trusted. All the truncations are
equally inadequate there, since the terms of the expansion never decrease.
This means that the problems become nonperturbative, when we approach the
peak.

In the case of fakeons, these facts point to a new type of uncertainty
relation, a \textquotedblleft peak uncertainty\textquotedblright\ $\Delta
E\gtrsim \Delta E_{\text{min}}=\Gamma /2$, which constrains our
observations. Knowing the very nature of the fakeon, it is not surprising
that we cannot approach its peak too closely, since, by definition, a purely
virtual particle refuses to be brought to reality.

Ultimately, the quantization prescription cannot be washed away by the
resummation. There is always a region where the true nature of the particle
we are studying (physical, fake or ghost) makes the difference.

The regime $\Delta E\gg \Gamma /2$ applies to quantum gravity and situations
like the one of the muon, where the resummation is unnecessary. The regime $%
\Delta E\ll \Gamma /2$ applies to collider physics and situations like the
one of the $Z$ boson. We have provided phenomenological candidates for the
dressed propagators close to the peaks, to describe what the nonperturbative
effects might look like there.

The peak uncertainty has no direct relation with the violation of
microcausality, also due to fakeons.

\vskip15truept \noindent {\large \textbf{Acknowledgments}}

\vskip 3truept

We are grateful to U. Aglietti, D. Comelli, E. Gabrielli, L. Marzola, M.
Piva and M. Raidal for helpful discussions. We thank the National Institute
of Chemical Physics and Biophysics (NICPB), Tallinn, Estonia, for
hospitality during the early stage of this work.

\vskip 1.5truecm

\noindent {\textbf{\huge Appendices}} \renewcommand{\thesection}{%
\Alph{section}} \renewcommand{\theequation}{\thesection.\arabic{equation}} %
\setcounter{section}{0}

\section{Alternative formal resummation for purely virtual particles}

\label{alternative}\setcounter{equation}{0}

In this appendix we derive the alternative formula (\ref{phatf2}) for the
formal dressed propagator of a purely virtual particle. The crucial point is
the role of the infinitesimal width $\epsilon $. Consider the identities%
\begin{eqnarray}
\sum_{n=0}^{\infty }(a+b)^{n} &=&\frac{1}{1-a}\sum_{n=0}^{\infty }\left( 
\frac{b}{1-a}\right) ^{n}=\frac{1}{1-a-b},  \notag \\
\sum_{n=0}^{\infty }(a+b+c)^{n} &=&\frac{1}{1-a-b}\sum_{n=0}^{\infty }\left( 
\frac{c}{1-a-b}\right) ^{n}=\frac{1}{1-a-b-c},  \label{identities}
\end{eqnarray}%
which hold perturbatively in $a$, $b$ and $c$. If we apply them to the sum
that appears in (\ref{sumfake}), with 
\begin{equation*}
a=-i(1-Z^{-1})(p^{2}-m^{2})P_{\chi },\quad b=-iZ^{-1}\Delta m^{2}P_{\chi
},\quad c=-Z^{-1}m_{\text{ph}}\Gamma P_{\chi },
\end{equation*}%
we can first sum the powers of $1-Z^{-1}$, then the powers of $\Delta m^{2}$
and finally the powers of $\Gamma $. This gives us the possibility to treat
them differently, because they are not on an equal footing with respect to
the problems that we discuss in the paper.

For example, the powers of $1-Z^{-1}$ can be summed with no difficulty,
because $Z$ is just the overall normalization. They give%
\begin{equation*}
P_{\chi }(p^{2},m^{2},\epsilon )\frac{1}{1-a}=\frac{iZ(p^{2}-m^{2})}{%
(p^{2}-m^{2})^{2}+\tilde{\epsilon}^{2}}=ZP_{\chi }(p^{2},m^{2},\tilde{%
\epsilon})\underset{\tilde{\epsilon}\rightarrow 0}{\rightarrow }\mathcal{P}%
\frac{iZ}{p^{2}-m^{2}}.
\end{equation*}%
In this derivation it does not really matter whether we remove $\epsilon $
term by term or after the sum.

If we proceed with the first line of (\ref{identities}), we need to calculate%
\begin{equation*}
P_{\chi }(p^{2},m^{2},\epsilon )\frac{1}{1-a}\sum_{n=0}^{\infty }\frac{b^{n}%
}{(1-a)^{n}}=ZP_{\chi }(p^{2},m^{2},\tilde{\epsilon})\sum_{n=0}^{\infty }%
\left[ -i\Delta m^{2}P_{\chi }(p^{2},m^{2},\tilde{\epsilon})\right] ^{n}.
\end{equation*}%
Taking $\epsilon $ to zero term-by-term, we get powers of the Cauchy
principal value. The coincidence-splitting method (see \cite{diagrammarMio}
and appendix \ref{split}) amounts to define such powers by starting from non
coincident singularities and using the identity (\ref{csplit}). Then we find%
\begin{equation}
iZ\sum_{n=0}^{\infty }\frac{(-\Delta m^{2})^{n}}{n!}\mathcal{P}^{(n)}\frac{1%
}{p^{2}-m^{2}}=\mathcal{P}\frac{iZ}{p^{2}-m_{\text{ph}}^{2}}\underset{\tilde{%
\epsilon}\rightarrow 0}{\leftarrow }\frac{iZ(p^{2}-m_{\text{ph}}^{2})}{%
(p^{2}-m_{\text{ph}}^{2})^{2}+\tilde{\epsilon}^{2}}=ZP_{\chi }(p^{2},m_{%
\text{ph}}^{2},\tilde{\epsilon})  \label{phdressed}
\end{equation}%
where $\mathcal{P}^{(n)}$ is the $n$-th derivative of the principal value.
We have restored $\tilde{\epsilon}$ in the last two expressions.

As far as the powers of $\Gamma $ are concerned, we cannot resum them with $%
\tilde{\epsilon}=0$, as shown in subsection \ref{anyway}. If we use the
second line of (\ref{identities}), we obtain%
\begin{equation*}
\hat{P}_{\chi }\simeq ZP_{\chi }(p^{2},m_{\text{ph}}^{2},\tilde{\epsilon}%
)\sum_{n=0}^{\infty }\left[ -m_{\text{ph}}\Gamma P_{\chi }(p^{2},m_{\text{ph}%
}^{2},\tilde{\epsilon})\right] ^{n}
\end{equation*}%
and the sum gives formula (\ref{phatf2}).

\section{Singular distributions}

\label{singulard}\setcounter{equation}{0}

A distribution is a continuous linear functional on the space of test
functions, which are the infinitely differentiable functions with compact
support. In this appendix we show that $\Delta _{\hat{\Gamma}}(x)$, defined
in formula (\ref{dgn}), is not a distribution. Let us start from the
truncated sums%
\begin{equation*}
\Delta _{\hat{\Gamma}}^{N}(x)\equiv \sum_{n=0}^{N}\frac{(-\hat{\Gamma}%
^{2})^{n}}{(2n)!}\delta ^{(2n)}(x)
\end{equation*}%
and check their actions on the function $g(x)=1/(x^{2}+\hat{\Gamma}^{2})$.
Since 
\begin{equation*}
\Delta _{\hat{\Gamma}}^{N}g\equiv \int_{-\infty }^{+\infty }\mathrm{d}x%
\hspace{0.01in}\Delta _{\hat{\Gamma}}^{N}(x)g(x)=\frac{N+1}{\hat{\Gamma}^{2}}%
,
\end{equation*}%
the limit $N\rightarrow \infty $ does not converge. We reach the same
conclusion on the test function%
\begin{equation*}
\hat{g}(x)=\left\{ 
\begin{tabular}{l}
$g(x)\mathrm{e}^{-\frac{R^{2}}{(R^{2}-x^{2})}}$ for $|x|<R,$ \\ 
0 for $|x|\geqslant R,$%
\end{tabular}%
\right.
\end{equation*}%
with $\hat{\Gamma}<R$, where $R>0$ is some radius. This proves that the
sequence of distributions $\Delta _{\hat{\Gamma}}^{N}$ does not converge to
a distribution.

Now, consider the functions 
\begin{equation}
f_{n}(x)=\frac{x-\frac{1}{n}}{\left( x-\frac{1}{n}\right) ^{2}+\hat{\Gamma}%
^{2}}.  \label{fin}
\end{equation}%
The sum of (\ref{dgn})\ converges for $\hat{\Gamma}<1/(2n)$, where it gives%
\begin{equation}
\Delta _{\hat{\Gamma}}f_{n}\equiv \int_{-\infty }^{+\infty }\mathrm{d}x%
\hspace{0.01in}\Delta _{\hat{\Gamma}}(x)f_{n}(x)\equiv -\frac{n(1+2n^{2}\hat{%
\Gamma}^{2})}{1+4n^{2}\hat{\Gamma}^{2}}.  \label{Dn}
\end{equation}%
It does not converge for $\hat{\Gamma}\geqslant 1/(2n)$. If, given $n$, we
define $\Delta _{\hat{\Gamma}}f_{n}$ to be the right-hand side of (\ref{Dn})
everywhere, by analytic continuation from the region $0<\hat{\Gamma}<1/(2n)$%
,\ we find another problem: $f_{n}(x)$ tends to $\hspace{0.01in}f_{\infty
}(x)=x/(x^{2}+\hat{\Gamma}^{2})$ for $n\rightarrow \infty $, but the
right-hand side of (\ref{Dn}) explodes.

The functions (\ref{fin}) are not test functions, but the test functions%
\begin{equation*}
\hat{f}_{n}(x)=\left\{ 
\begin{tabular}{l}
$f_{n}(x)\mathrm{e}^{-\frac{R^{2}}{(R^{2}-x^{2})}}$ for $|x|<R,$ \\ 
0 for $|x|\geqslant R,$%
\end{tabular}%
\right.
\end{equation*}%
lead to the same conclusions for $\hat{\Gamma}<R$. Precisely, $\Delta _{\hat{%
\Gamma}}\hat{f}_{n}$ is equal to the right-hand side of (\ref{Dn}) times $%
\mathrm{\exp }(-R^{2}/(R^{2}+\hat{\Gamma}^{2}))$, so the product still blows
up for $n\rightarrow \infty $. This means, again, that $\Delta _{\hat{\Gamma}%
}(x)$ is not a continuous linear functional, i.e., not a distribution.

\section{Coincidence-splitting method}

\label{split}\setcounter{equation}{0}

In this appendix we explain how to define the products of principal values
and delta functions by means of the coincidence-splitting method or, when
necessary, its symmetrized version. The idea is to treat coincident
singularities as the limits of distinct ones.

First observe that the powers of $\delta (x)$ higher than one are set to
zero by this method. The powers of the Cauchy principal value were
considered in \cite{FLRW}, where it was proved that%
\begin{equation}
\lim_{\delta _{i}\rightarrow 0}\mathcal{P}\prod\limits_{i=1}^{n+1}\frac{1}{%
x-\delta _{i}}=\frac{(-1)^{n}}{n!}\mathcal{P}^{(n)}\frac{1}{x},
\label{csplit}
\end{equation}%
(the parameters $\delta _{i}$ being different from one another).

The product of one delta function times powers of the principal value gives

\begin{equation}
{\text{s}}\!\!\lim_{\delta _{i}\rightarrow 0}\delta (x-\delta _{1})\mathcal{P%
}\prod\limits_{i=2}^{n+1}\frac{1}{x-\delta _{i}}=\frac{(-1)^{n}}{(n+1)!}%
\delta ^{(n)}(x),  \label{csplit2}
\end{equation}%
where \textquotedblleft slim\textquotedblright\ means that the expression
must be symmetrized in $\delta _{i}$, $i=1,\ldots n+1$ before taking the
limit. The identity (\ref{csplit2}) can be proved by acting on a test
function $f(x)$. We write%
\begin{equation*}
{\text{s}}\!\!\lim_{\delta _{i}\rightarrow 0}\int_{-\infty }^{+\infty }%
\mathrm{d}x\hspace{0.01in}f(x)\delta (x-\delta _{1})\mathcal{P}%
\prod\limits_{i=2}^{n+1}\frac{1}{x-\delta _{i}}=\frac{1}{n+1}\lim_{\delta
_{i}\rightarrow 0}\sum_{j=1}^{n+1}f(\delta _{j})\prod\limits_{i=1,i\neq
j}^{n+1}\frac{1}{\delta _{j}-\delta _{i}},
\end{equation*}%
then Taylor expand $f(x)$ around zero and finally use the formula%
\begin{equation*}
\sum_{j=1}^{n+1}\delta _{j}^{k}\prod\limits_{i=1,i\neq j}^{n+1}\frac{1}{%
\delta _{j}-\delta _{i}}=\left\{ 
\begin{tabular}{l}
$0\qquad $for $k<n$ \\ 
$1\qquad $for $k=n$ \\ 
polynomial of $\delta _{j}$ for $k>n.$%
\end{tabular}%
\right.
\end{equation*}%
The result is $f^{(n)}(0)/(n+1)!$, in agreement with (\ref{csplit2}).

We can use these results, for example, to study the expansions of $\Omega
_{\varphi \hspace{0.01in}\text{particle}}$ and $\Omega _{\varphi \hspace{%
0.01in}\text{decay}}$ by letting $\epsilon $ tend to zero term by term. The
diagrams listed in the first and second lines of fig. \ref{cutdiagrams} give%
\begin{eqnarray}
2\Omega _{\varphi \hspace{0.01in}\text{particle}} &=&\sum_{k=0}^{\infty
}(i\Sigma ^{\ast }P_{\varphi }^{\ast })^{k}(P_{\varphi }+P_{\varphi }^{\ast
})\sum_{n=0}^{\infty }(-i\Sigma P_{\varphi })^{n},  \notag \\
2\Omega _{\varphi \hspace{0.01in}\text{decay}} &=&P_{\varphi }^{\ast
}\sum_{k=0}^{\infty }(i\Sigma ^{\ast }P_{\varphi }^{\ast })^{k}(-2\text{Im}%
[\Sigma ])\sum_{n=0}^{\infty }(-i\Sigma P_{\varphi })^{n}P_{\varphi },
\label{omica4}
\end{eqnarray}%
where $P_{\varphi }+P_{\varphi }^{\ast }$ is the cut propagator and $2$Im$%
[\Sigma ]=-i\Sigma +(-i\Sigma )^{\ast }$ is the cut bubble diagram. We apply
the approximation (\ref{approx}) with $\Delta m^{2}=0$, $Z=1$, to focus on
the powers of $\Gamma $. Using $m^{2}P_{\varphi }=\pi \delta (x)+i\mathcal{P}%
(1/x)$, $P_{\varphi }+P_{\varphi }^{\ast }=$ $2\pi \delta (x)/m^{2}$ and $2$%
Im$[\Sigma ]=-2m^{2}\hat{\Gamma}$, where $x=(p^{2}-m^{2})/m^{2}$ and $\hat{%
\Gamma}=\Gamma /m$, and dealing with the products of delta functions and
principal values by means of formulas (\ref{csplit})\ and (\ref{csplit2}),
we obtain 
\begin{eqnarray}
\Omega _{\varphi \hspace{0.01in}\text{particle}} &\rightarrow &\frac{\pi }{%
m^{2}}\sum_{n=0}^{\infty }\frac{(-\hat{\Gamma}^{2})^{n}}{(2n+1)!}\delta
^{(2n)}(x),  \notag \\
\Omega _{\varphi \hspace{0.01in}\text{decay}} &\rightarrow &\frac{1}{m^{2}%
\hat{\Gamma}}\sum_{n=0}^{\infty }\frac{(-\hat{\Gamma}^{2})^{n+1}}{(2n+1)!}%
\mathcal{P}^{(2n+1)}\frac{1}{x}+\frac{\pi }{m^{2}}\sum_{n=0}^{\infty }\frac{%
2n(-\hat{\Gamma}^{2})^{n}}{(2n+1)!}\delta ^{(2n)}(x).  \label{omica3}
\end{eqnarray}%
Taken separately, these expressions are not well-defined distributions.
However, their sum is. Indeed, formula (\ref{physum}) gives%
\begin{equation}
\Omega _{\varphi \hspace{0.01in}\text{particle}}+\Omega _{\varphi \hspace{%
0.01in}\text{decay}}\rightarrow \frac{1}{m^{2}}\frac{\hat{\Gamma}}{x^{2}+%
\hat{\Gamma}^{2}},  \label{bw}
\end{equation}%
which is the expected Breit-Wigner function.

\end{document}